\documentclass[a4paper,11pt]{article}
\usepackage{jheppub} 
\usepackage[T1]{fontenc} 
\usepackage{amsmath,amsthm}
\usepackage{amssymb, amsfonts}
\usepackage{bbm}

\usepackage{dcolumn}
\usepackage{bm}

\usepackage[T1]{fontenc} 
\usepackage{amsmath,amsthm}
\usepackage{amssymb, amsfonts}
\usepackage{bbm}
\usepackage{graphicx}
\usepackage{dsfont}
\usepackage{relsize}
\usepackage{stmaryrd}
\usepackage{lmodern}
\usepackage{slantsc}
\usepackage{scalefnt}
\usepackage{wasysym}
\usepackage{xspace}
\usepackage{boxedminipage}
\usepackage{ifpdf}
\usepackage{multirow, hhline}
\usepackage{slashed}
\usepackage{xifthen}
\usepackage{tensor}
\usepackage{array, nicematrix}
\usepackage{tabu}
\usepackage{pbox}
\usepackage{multirow}
\usepackage{tikz}
\usepackage{overpic}
\usepackage{color}
\usepackage{diagbox}
\usepackage{makecell}
\usetikzlibrary{arrows, matrix, calc, scopes, decorations.markings}
\usepackage[mathcal]{euscript}
\usepackage{booktabs}
\usepackage{autobreak, subfig, diagbox, mathrsfs}
\usepackage{mathbbol}
\usepackage{tcolorbox}
\usepackage{hyperref}
\usepackage{adjustbox}
\allowdisplaybreaks[4]

\newcommand{\NN}{\mathbb{N}}
\newcommand{\Z}{\mathbb{Z}}

\newcommand{\M}{\mathcal{M}}
\newcommand{\Hil}{\mathcal{H}}

\newcommand{\Cent}{\mathcal{Z}}
\newcommand{\T}{\mathcal{T}}

\newcommand{\C}{\mathbb{C}}

\newcommand{\Fus}{\mathscr{F}}

\newcommand{\RR}{\mathbb{R}}
\newcommand{\CC}{\mathbb{C}}
\newcommand{\idm}{{\mathds{1}}}

\newcommand{\df}{{\mathrm{d}}}

\newcommand{\dk}[2][1]{{\ifthenelse{\equal{#1}{1}}{\frac{\df{#2}}{2\pi}}{\frac{\df^{#1}{#2}}{(2\pi)^{#1}}}}}

\renewcommand{\Vec}{{\rm Vec}}

\newcommand{\ket}[1]{{\left\vert{#1}\right\rangle}}

\newcommand{\eq}[1]{\begin{align*}#1\end{align*}}
\newcommand{\eqn}[2][0]{\ifthenelse{\equal{#1}{0}}{\begin{equation}\begin{aligned}#2\end{aligned}\end{equation}}{\begin{equation}\begin{aligned}#2\end{aligned}\label{#1}\end{equation}}}

\usetikzlibrary{shapes.geometric, snakes}
\tikzset{>=latex}
\usetikzlibrary{arrows, matrix, calc, scopes, decorations.markings}
\usetikzlibrary{decorations.pathmorphing, positioning}
\tikzset{snake it/.style={decorate, decoration={snake,amplitude=0.2mm,segment length=1mm}}}
\tikzset{->-/.style={decoration={
			 markings,
			 mark=at position .5*\pgfdecoratedpathlength+2pt with {\arrow{>}}},postaction={decorate}}}

\tikzset{-<-/.style={decoration={
			 markings,
			 mark=at position .5*\pgfdecoratedpathlength+2pt with {\arrow{<}}},postaction={decorate}}}

\newtabulinestyle{dash=off 2pt}

{\null\,\vcenter\bgroup\scriptsize
\arraycolsep=.13885em
\hbox\bgroup$\array{@{}#1@{}}}
{\endarray$\egroup\egroup\,\null}
\newcommand{\Lattice}{
\begin{tikzpicture}[baseline={([yshift=-.8ex]current bounding box.center)},scale=1]
\draw [decorate,decoration={snake,segment length=.12cm, amplitude=.04cm}] (1.7,6.8) -- (2.3,6.8);
\draw [] (1.7,6.4) -- (1.7,6.8);
\draw [] (1.7,6.8) -- (1.7,7.2);
\draw [] (1.7,7.2) -- (2.4,7.6);
\draw [] (2.4,7.6) -- (3.1,7.2);
\draw [] (2.4,6.) -- (1.7,6.4);
\draw [] (3.1,6.4) -- (2.4,6.);
\draw [] (2.4,5.2) -- (2.4,5.6);
\draw [] (2.4,5.6) -- (2.4,6.);
\draw [decorate,decoration={snake,segment length=.12cm, amplitude=.04cm}] (2.4,5.6) -- (3.,5.6);
\draw [decorate,decoration={snake,segment length=.12cm, amplitude=.04cm}] (3.1,6.8) -- (3.7,6.8);
\draw [] (3.1,6.4) -- (3.1,6.8);
\draw [] (3.1,6.8) -- (3.1,7.2);
\draw [] (3.1,7.2) -- (3.8,7.6);
\draw [] (3.8,7.6) -- (4.5,7.2);
\draw [] (3.8,6.) -- (3.1,6.4);
\draw [] (4.5,6.4) -- (3.8,6.);
\draw [] (3.8,5.2) -- (3.8,5.6);
\draw [] (3.8,5.6) -- (3.8,6.);
\draw [decorate,decoration={snake,segment length=.12cm, amplitude=.04cm}] (3.8,5.6) -- (4.4,5.6);
\draw [decorate,decoration={snake,segment length=.12cm, amplitude=.04cm}] (4.5,6.8) -- (5.1,6.8);
\draw [] (4.5,6.4) -- (4.5,6.8);
\draw [] (4.5,6.8) -- (4.5,7.2);
\draw [] (4.5,7.2) -- (5.2,7.6);
\draw [] (5.2,7.6) -- (5.9,7.2);
\draw [] (5.2,6.) -- (4.5,6.4);
\draw [] (5.9,6.4) -- (5.2,6.);
\draw [] (5.2,5.2) -- (5.2,5.6);
\draw [] (5.2,5.6) -- (5.2,6.);
\draw [decorate,decoration={snake,segment length=.12cm, amplitude=.04cm}] (5.2,5.6) -- (5.8,5.6);
\draw [decorate,decoration={snake,segment length=.12cm, amplitude=.04cm}] (5.9,6.8) -- (6.5,6.8);
\draw [] (5.9,6.4) -- (5.9,6.8);
\draw [] (5.9,6.8) -- (5.9,7.2);
\draw [] (5.9,7.2) -- (6.6,7.6);
\draw [] (6.6,7.6) -- (7.3,7.2);
\draw [] (6.6,6.) -- (5.9,6.4);
\draw [] (7.3,6.4) -- (6.6,6.);
\draw [] (6.6,5.2) -- (6.6,5.6);
\draw [] (6.6,5.6) -- (6.6,6.);
\draw [decorate,decoration={snake,segment length=.12cm, amplitude=.04cm}] (6.6,5.6) -- (7.2,5.6);
\draw [decorate,decoration={snake,segment length=.12cm, amplitude=.04cm}] (1.7,4.4) -- (2.3,4.4);
\draw [] (1.7,4.) -- (1.7,4.4);
\draw [] (1.7,4.4) -- (1.7,4.8);
\draw [] (1.7,4.8) -- (2.4,5.2);
\draw [] (2.4,5.2) -- (3.1,4.8);
\draw [] (2.4,3.6) -- (1.7,4.);
\draw [] (3.1,4.) -- (2.4,3.6);
\draw [] (2.4,3.2) -- (2.4,3.6);
\draw [decorate,decoration={snake,segment length=.12cm, amplitude=.04cm}] (3.1,4.4) -- (3.7,4.4);
\draw [] (3.1,4.) -- (3.1,4.4);
\draw [] (3.1,4.4) -- (3.1,4.8);
\draw [] (3.1,4.8) -- (3.8,5.2);
\draw [] (3.8,5.2) -- (4.5,4.8);
\draw [] (3.8,3.6) -- (3.1,4.);
\draw [] (4.5,4.) -- (3.8,3.6);
\draw [] (3.8,3.2) -- (3.8,3.6);
\draw [decorate,decoration={snake,segment length=.12cm, amplitude=.04cm}] (4.5,4.4) -- (5.1,4.4);
\draw [] (4.5,4.) -- (4.5,4.4);
\draw [] (4.5,4.4) -- (4.5,4.8);
\draw [] (4.5,4.8) -- (5.2,5.2);
\draw [] (5.2,5.2) -- (5.9,4.8);
\draw [] (5.2,3.6) -- (4.5,4.);
\draw [] (5.9,4.) -- (5.2,3.6);
\draw [] (5.2,3.2) -- (5.2,3.6);
\draw [decorate,decoration={snake,segment length=.12cm, amplitude=.04cm}] (5.9,4.4) -- (6.5,4.4);
\draw [] (5.9,4.) -- (5.9,4.4);
\draw [] (5.9,4.4) -- (5.9,4.8);
\draw [] (5.9,4.8) -- (6.6,5.2);
\draw [] (6.6,5.2) -- (7.3,4.8);
\draw [] (6.6,3.6) -- (5.9,4.);
\draw [] (7.3,4.) -- (6.6,3.6);
\draw [] (6.6,3.2) -- (6.6,3.6);
\draw [decorate,decoration={snake,segment length=.12cm, amplitude=.04cm}] (7.3,6.8) -- (7.9,6.8);
\draw [] (7.3,6.4) -- (7.3,6.8);
\draw [] (7.3,6.8) -- (7.3,7.2);
\draw [] (7.3,7.2) -- (7.6,7.4);
\draw [] (8.,6.) -- (7.3,6.4);
\draw [] (8.3,6.2) -- (8.,6.);
\draw [] (8.,5.2) -- (8.,5.6);
\draw [] (8.,5.6) -- (8.,6.);
\draw [decorate,decoration={snake,segment length=.12cm, amplitude=.04cm}] (8.,5.6) -- (8.6,5.6);
\draw [decorate,decoration={snake,segment length=.12cm, amplitude=.04cm}] (7.3,4.4) -- (7.9,4.4);
\draw [] (7.3,4.) -- (7.3,4.4);
\draw [] (7.3,4.4) -- (7.3,4.8);
\draw [] (7.3,4.8) -- (8.,5.2);
\draw [] (8.,5.2) -- (8.3,5.);
\draw [] (8.,3.6) -- (7.3,4.);
\draw [] (8.,3.2) -- (8.,3.6);
\draw [] (6.6,7.6) -- (6.6,8.);
\draw [] (5.2,7.6) -- (5.2,8.);
\draw [] (3.8,7.6) -- (3.8,8.);
\draw [] (2.4,7.6) -- (2.4,8.);
\draw [] (8.3,3.8) -- (8.,3.6);
\draw [] (1.7,4.) -- (1.3,3.8);
\draw [] (1.7,4.8) -- (1.3,5.);
\draw [] (1.7,6.4) -- (1.3,6.2);
\draw [] (1.7,7.2) -- (1.3,7.4);
\end{tikzpicture}
}

\newcommand{\PlaquetteSrc}{
\begin{tikzpicture}[baseline={([yshift=-.8ex]current bounding box.center)},scale=.5]
\draw [->-] (3.2,5.6) -- (3.2,6.8);
\draw [->-] (3.2,7.2) -- (4.6,8.);
\draw [->-] (4.6,8.) -- (6.,7.2);
\draw [->-] (6.,5.6) -- (4.6,4.8);
\draw [->-] (4.6,4.8) -- (3.2,5.6);
\draw [->-] (6.,7.2) -- (6.,5.6);
\draw [decorate,decoration={snake,segment length=.12cm, amplitude=.04cm}, ->] (3.2,6.4) -- (4.4,6.4);
\draw [->-] (2.6,7.6) -- (3.2,7.2);
\draw [->-] (4.6,8.8) -- (4.6,8.);
\draw [->-] (6.6,7.6) -- (6.,7.2);
\draw [->-] (6.6,5.2) -- (6.,5.6);
\draw [->-] (4.6,4.) -- (4.6,4.8);
\draw [->-] (2.6,5.2) -- (3.2,5.6);
\draw [->-] (3.2,6.6) -- (3.2,7.2);
\node [centered, rotate=0.] at (2.4,7.8) {\scriptsize $e_1$};
\node [centered, rotate=0.] at (4.95,8.8) {\scriptsize $e_2$};
\node [centered, rotate=0.] at (6.9,7.8) {\scriptsize $e_3$};
\node [centered, rotate=0.] at (6.9,5.) {\scriptsize $e_4$};
\node [centered, rotate=0.] at (5.,4.) {\scriptsize $e_5$};
\node [centered, rotate=0.] at (2.3,5.) {\scriptsize $e_6$};
\node [centered, rotate=0.] at (2.85,6.) {\scriptsize $i_7$};
\node [centered, rotate=0.] at (2.85,6.9) {\scriptsize $i_1$};
\node [centered, rotate=0.] at (3.6,7.9) {\scriptsize $i_2$};
\node [centered, rotate=0.] at (5.6,7.9) {\scriptsize $i_3$};
\node [centered, rotate=0.] at (6.4,6.4) {\scriptsize $i_4$};
\node [centered, rotate=0.] at (5.6,4.9) {\scriptsize $i_5$};
\node [centered, rotate=0.] at (3.7,4.9) {\scriptsize $i_6$};
\node [centered, rotate=0.] at (4.7,6.4) {\scriptsize $p$};
\end{tikzpicture}
}

\newcommand{\PlaquetteTar}{
\begin{tikzpicture}[baseline={([yshift=-.8ex]current bounding box.center)},scale=.5]
\draw [->-] (3.2,5.6) -- (3.2,6.8);
\draw [->-] (3.2,7.2) -- (4.6,8.);
\draw [->-] (4.6,8.) -- (6.,7.2);
\draw [->-] (6.,5.6) -- (4.6,4.8);
\draw [->-] (4.6,4.8) -- (3.2,5.6);
\draw [->-] (6.,7.2) -- (6.,5.6);
\draw [decorate,decoration={snake,segment length=.12cm, amplitude=.04cm}] (3.2,6.4) -- (4.4,6.4);
\draw [->-] (2.6,7.6) -- (3.2,7.2);
\draw [->-] (4.6,8.8) -- (4.6,8.);
\draw [->-] (6.6,7.6) -- (6.,7.2);
\draw [->-] (6.6,5.2) -- (6.,5.6);
\draw [->-] (4.6,4.) -- (4.6,4.8);
\draw [->-] (2.6,5.2) -- (3.2,5.6);
\draw [->-] (3.2,6.6) -- (3.2,7.2);
\node [centered, rotate=0.] at (2.4,7.8) {\scriptsize $e_1$};
\node [centered, rotate=0.] at (4.95,8.8) {\scriptsize $e_2$};
\node [centered, rotate=0.] at (6.9,7.8) {\scriptsize $e_3$};
\node [centered, rotate=0.] at (6.9,5.) {\scriptsize $e_4$};
\node [centered, rotate=0.] at (5.,4.) {\scriptsize $e_5$};
\node [centered, rotate=0.] at (2.3,5.) {\scriptsize $e_6$};
\node [centered, rotate=0.] at (2.85,6.) {\scriptsize $j_7$};
\node [centered, rotate=0.] at (2.85,6.9) {\scriptsize $j_1$};
\node [centered, rotate=0.] at (3.6,7.9) {\scriptsize $j_2$};
\node [centered, rotate=0.] at (5.6,7.9) {\scriptsize $j_3$};
\node [centered, rotate=0.] at (6.4,6.4) {\scriptsize $j_4$};
\node [centered, rotate=0.] at (5.6,4.9) {\scriptsize $j_5$};
\node [centered, rotate=0.] at (3.7,4.9) {\scriptsize $j_6$};
\node [centered, rotate=0.] at (4.7,6.4) {\scriptsize $1$};
\end{tikzpicture}
}

\newcommand{\PachnerOne}{\begin{tikzpicture}[baseline={([yshift=-.8ex]current bounding box.center)},scale=.6]
\draw [->-] (0.,0.2) -- (.5,1.);
\draw [->-] (0.,1.8) -- (.5,1.);
\draw [->-] (2.,1.) -- (.5,1.);
\draw [->-] (2.5,1.8) -- (2.,1.);
\draw [->-] (2.5,0.2) -- (2.,1.);
\node at (1.25, 1.3) {\scriptsize $m$};
\node at (-.15, 2.) {\scriptsize $a$};
\node at (-.15, 0.) {\scriptsize $b$};
\node at (2.65, 2.) {\scriptsize $d$};
\node at (2.65, 0.) {\scriptsize $c$};
\end{tikzpicture}\qquad = \qquad \sum_{n\in L_{\Fus}}\ \sqrt{d_md_n}\ G^{abm}_{cdn}\quad \begin{tikzpicture}[baseline={([yshift=-.8ex]current bounding box.center)},scale=.6]
\draw [->-] (0.,0.) -- (.8,.5);
\draw [->-] (1.6,0.) -- (.8,.5);
\draw [->-] (.8,.5) -- (.8,2.);
\draw [->-] (0.,2.5) -- (.8,2.);
\draw [->-] (1.6,2.5) -- (.8,2.);
\node at (1.05, 1.3) {\scriptsize $n$};
\node at (-.15, 2.6) {\scriptsize $a$};
\node at (-.15, -.1) {\scriptsize $b$};
\node at (1.75, 2.6) {\scriptsize $d$};
\node at (1.75, -0.1) {\scriptsize $c$};
\end{tikzpicture}}

\newcommand{\PachnerTwo}{\begin{tikzpicture}[baseline={([yshift=-.8ex]current bounding box.center)},scale=.6]
\draw [->-] (1, 0) -- (1, 1);
\draw [->-] (1, 1) arc (270:90:.5);
\draw [->-] (1, 1) arc (-90:90:.5);
\draw [->-] (1, 2) -- (1, 3);
\node at (1.25, .3) {\scriptsize $a$};
\node at (1., 1.5) {\scriptsize \color{red}$\times$};
\node at (.2, 1.5) {\scriptsize $x$};
\node at (1.8, 1.5) {\scriptsize $y$};
\node at (1.25, 2.7) {\scriptsize $b$};
\end{tikzpicture}\qquad = \qquad \sqrt{\frac{d_xd_y}{d_a}}\ \delta_{ab}\ \delta_{xya^\ast}\quad \begin{tikzpicture}[baseline={([yshift=-.8ex]current bounding box.center)},scale=.5]
\draw [->-] (1, 0) -- (1, 3);
\node at (1.25, 1.5) {\scriptsize $a$};
\end{tikzpicture}}

\newcommand{\PachnerThree}{\begin{tikzpicture}[baseline={([yshift=-.8ex]current bounding box.center)},scale=.5]
\draw [->-] (1, 0) -- (1, 3);
\node at (1.25, 1.5) {\scriptsize $a$};
\end{tikzpicture}\quad = \quad \frac{1}{D}\sum_{xy\in L_{\Fus}} \sqrt{\frac{d_xd_y}{d_a}}\ \delta_{xya^\ast} \quad \begin{tikzpicture}[baseline={([yshift=-.8ex]current bounding box.center)},scale=.6]
\draw [->-] (1, 0) -- (1, 1);
\draw [->-] (1, 1) arc (270:90:.5);
\draw [->-] (1, 1) arc (-90:90:.5);
\draw [->-] (1, 2) -- (1, 3);
\node at (1.25, .3) {\scriptsize $a$};
\node at (.2, 1.5) {\scriptsize $x$};
\node at (1.8, 1.5) {\scriptsize $y$};
\node at (1.25, 2.7) {\scriptsize $a$};
\end{tikzpicture}}

\newcommand{\PachnerFour}{
\begin{tikzpicture}[baseline={([yshift=-.8ex]current bounding box.center)},scale=.6]
\draw [->-] (3.2,5.6) -- (3.2,6.8);
\draw [->-] (3.2,7.2) -- (4.6,8.);
\draw [->-] (4.6,8.) -- (6.,7.2);
\draw [->-] (6.,5.6) -- (4.6,4.8);
\draw [->-] (4.6,4.8) -- (3.2,5.6);
\draw [->-] (6.,7.2) -- (6.,5.6);
\draw [decorate,decoration={snake,segment length=.12cm, amplitude=.04cm}, ->] (3.2,6.6) -- (4.4,6.6);
\draw [->-] (2.6,7.6) -- (3.2,7.2);
\draw [->-] (4.6,8.8) -- (4.6,8.);
\draw [->-] (6.6,7.6) -- (6.,7.2);
\draw [->-] (6.6,5.2) -- (6.,5.6);
\draw [->-] (4.6,4.) -- (4.6,4.8);
\draw [->-] (2.6,5.2) -- (3.2,5.6);
\draw [->-] (3.2,6.6) -- (3.2,7.2);
\node [centered, rotate=0.] at (2.5,7.8) {\scriptsize $e_1$};
\node [centered, rotate=0.] at (2.9,6.) {\scriptsize $i_0$};
\node [centered, rotate=0.] at (2.9,6.9) {\scriptsize $i_1$};
\node [centered, rotate=0.] at (3.6,7.8) {\scriptsize $i_2$};
\node [centered, rotate=0.] at (4.7,6.6) {\scriptsize $p$};
\end{tikzpicture}\ =\quad\sum_{j\in L_\Fus}\sqrt{d_{i_1}d_j}\ G^{i_2^\ast e_1i_1}_{i_0 p^\ast j}\ 
\begin{tikzpicture}[baseline={([yshift=-.8ex]current bounding box.center)},scale=.6]
\draw [->-] (3.2,5.6) -- (3.2,7.2);
\draw [->-] (3.2,7.2) -- (3.55,7.4);
\draw [->-] (3.55,7.4) -- (4.6,8.);
\draw [->-] (4.6,8.) -- (6.,7.2);
\draw [->-] (6.,5.6) -- (4.6,4.8);
\draw [->-] (4.6,4.8) -- (3.2,5.6);
\draw [->-] (6.,7.2) -- (6.,5.6);
\draw [->-] (2.6,7.6) -- (3.2,7.2);
\draw [->-] (4.6,8.8) -- (4.6,8.);
\draw [->-] (6.6,7.6) -- (6.,7.2);
\draw [->-] (6.6,5.2) -- (6.,5.6);
\draw [->-] (4.6,4.) -- (4.6,4.8);
\draw [->-] (2.6,5.2) -- (3.2,5.6);
\draw [decorate,decoration={snake,segment length=.12cm, amplitude=.04cm}, ->] (3.59,7.42) -- (4.1,6.5);
\node [centered, rotate=0.] at (2.5,5.) {\scriptsize $e_n$};
\node [centered, rotate=0.] at (2.5,7.8) {\scriptsize $e_1$};
\node [centered, rotate=0.] at (2.9,6.5) {\scriptsize $i_0$};
\node [centered, rotate=0.] at (3.3,7.6) {\scriptsize $j$};
\node [centered, rotate=0.] at (3.9,8.) {\scriptsize $i_2$};
\node [centered, rotate=0.] at (4.3,6.3) {\scriptsize $p$};
\node [centered, rotate=0.] at (3.7,5.1) {\scriptsize $i_n$};
\end{tikzpicture}
}

\newcommand{\PachnerFive}{
\begin{tikzpicture}[baseline={([yshift=-.8ex]current bounding box.center)},scale=.6]
\draw [->-] (3.2,5.6) -- (3.2,6.1);
\draw [->-] (3.2,6.1) -- (3.2,6.7);
\draw [->-] (3.2,6.7) -- (3.2,7.2);
\draw [->-] (3.2,7.2) -- (4.6,8.);
\draw [->-] (4.6,8.) -- (6.,7.2);
\draw [->-] (6.,5.6) -- (4.6,4.8);
\draw [->-] (4.6,4.8) -- (3.2,5.6);
\draw [->-] (6.,7.2) -- (6.,5.6);
\draw [decorate,decoration={snake,segment length=.12cm, amplitude=.04cm}, ->] (3.2,6.1) -- (4.4,6.1);
\draw [decorate,decoration={snake,segment length=.12cm, amplitude=.04cm}, ->] (3.2,6.7) -- (4.4,6.7);
\draw [->-] (2.6,7.6) -- (3.2,7.2);
\draw [->-] (4.6,8.8) -- (4.6,8.);
\draw [->-] (6.6,7.6) -- (6.,7.2);
\draw [->-] (6.6,5.2) -- (6.,5.6);
\draw [->-] (4.6,4.) -- (4.6,4.8);
\draw [->-] (2.6,5.2) -- (3.2,5.6);
\node [centered, rotate=0.] at (2.9,5.8) {\scriptsize $i_0$};
\node [centered, rotate=0.] at (2.9,7.1) {\scriptsize $i_1$};
\node [centered, rotate=0.] at (4.7,6.7) {\scriptsize $r$};
\node [centered, rotate=0.] at (4.7,6.1) {\scriptsize $s$};
\node [centered, rotate=0.] at (2.9,6.4) {\scriptsize $j$};
\end{tikzpicture}\quad = \quad \sum_{u\in L_\Fus}\sqrt{d_jd_p}\ G^{r^\ast i_1^\ast j}_{i_0s^\ast p}\quad 
\begin{tikzpicture}[baseline={([yshift=-.8ex]current bounding box.center)},scale=.6]
\draw [->-] (3.2,5.6) -- (3.2,6.8);
\draw [->-] (3.2,7.2) -- (4.6,8.);
\draw [->-] (4.6,8.) -- (6.,7.2);
\draw [->-] (6.,5.6) -- (4.6,4.8);
\draw [->-] (4.6,4.8) -- (3.2,5.6);
\draw [->-] (6.,7.2) -- (6.,5.6);
\draw [decorate,decoration={snake,segment length=.12cm, amplitude=.04cm}, ->] (3.2,6.6) -- (4.4,6.6);
\draw [->-] (2.6,7.6) -- (3.2,7.2);
\draw [->-] (4.6,8.8) -- (4.6,8.);
\draw [->-] (6.6,7.6) -- (6.,7.2);
\draw [->-] (6.6,5.2) -- (6.,5.6);
\draw [->-] (4.6,4.) -- (4.6,4.8);
\draw [->-] (2.6,5.2) -- (3.2,5.6);
\draw [->-] (3.2,6.6) -- (3.2,7.2);
\node [centered, rotate=0.] at (2.9,6.) {\scriptsize $i_0$};
\node [centered, rotate=0.] at (2.9,6.9) {\scriptsize $i_1$};
\node [centered, rotate=0.] at (4.7,6.6) {\scriptsize $p$};
\end{tikzpicture}
}

\newcommand{\PlaquetteMsr}{
\begin{tikzpicture}[baseline={([yshift=-.8ex]current bounding box.center)},scale=.55]
\draw [->-] (3.2,5.6) -- (3.2,6.8);
\draw [->-] (3.2,7.2) -- (4.6,8.);
\draw [->-] (4.6,8.) -- (6.,7.2);
\draw [->-] (6.,5.6) -- (4.6,4.8);
\draw [->-] (4.6,4.8) -- (3.2,5.6);
\draw [->-] (6.,7.2) -- (6.,5.6);
\draw [->-] (2.6,7.6) -- (3.2,7.2);
\draw [->-] (4.6,8.8) -- (4.6,8.);
\draw [->-] (6.6,7.6) -- (6.,7.2);
\draw [->-] (6.6,5.2) -- (6.,5.6);
\draw [->-] (4.6,4.) -- (4.6,4.8);
\draw [->-] (2.6,5.2) -- (3.2,5.6);
\draw [->-] (3.2,6.6) -- (3.2,7.2);
\node [centered, rotate=0.] at (2.4,7.8) {\scriptsize $e_1$};
\node [centered, rotate=0.] at (4.95,8.8) {\scriptsize $e_2$};
\node [centered, rotate=0.] at (6.9,7.8) {\scriptsize $e_3$};
\node [centered, rotate=0.] at (6.9,5.) {\scriptsize $e_4$};
\node [centered, rotate=0.] at (5.,4.) {\scriptsize $e_5$};
\node [centered, rotate=0.] at (2.3,5.) {\scriptsize $e_6$};
\node [centered, rotate=0.] at (2.85,6.) {\scriptsize $i_7$};
\node [centered, rotate=0.] at (2.85,6.9) {\scriptsize $i_1$};
\node [centered, rotate=0.] at (3.6,7.9) {\scriptsize $i_2$};
\node [centered, rotate=0.] at (5.6,7.9) {\scriptsize $i_3$};
\node [centered, rotate=0.] at (6.4,6.4) {\scriptsize $i_4$};
\node [centered, rotate=0.] at (5.6,4.9) {\scriptsize $i_5$};
\node [centered, rotate=0.] at (3.7,4.9) {\scriptsize $i_6$};

\draw [decorate,decoration={snake,segment length=.12cm, amplitude=.04cm}, ->] (3.2,6.4) -- (5.,6.4);
\draw [->] (3.8, 6.4) arc (180:0:.8);
\draw [] (4.4, 6.4) arc (-180:0:0.5);
\node [centered, rotate=0.] at (4.8,6.7) {\scriptsize $p$};
\node [centered, rotate=0.] at (4.1,6.1) {\scriptsize $t$};
\node [centered, rotate=0.] at (3.5,6.1) {\scriptsize $p$};
\node [centered, rotate=0.] at (5.65,6.4) {\scriptsize $s$};
\node [centered, rotate=0.] at (4.6,5.4) {\color{red} $\times$};
\end{tikzpicture}
}

\newcommand{\ExcitedA}{
\begin{tikzpicture}[baseline={([yshift=-.8ex]current bounding box.center)},scale=.7]
\draw [->-] (3.2,5.6) -- (3.2,7.2);
\node [centered, rotate=0.] at (2.9,6.4) {\scriptsize $j$};
\end{tikzpicture}
}

\newcommand{\ExcitedB}{
\begin{tikzpicture}[baseline={([yshift=-.8ex]current bounding box.center)},scale=.8]
\draw [->-] (3.2,5.6) -- (3.2,6.1);
\draw [->-] (3.2,6.1) -- (3.2,6.7);
\draw [->-] (3.2,6.7) -- (3.2,7.2);
\draw [decorate,decoration={snake,segment length=.12cm, amplitude=.04cm}, ->] (3.2,6.7) -- (4.,6.7);
\draw [decorate,decoration={snake,segment length=.12cm, amplitude=.04cm}, ->] (3.2,6.1) -- (2.4,6.1);
\node [centered, rotate=0.] at (3.,5.8) {\scriptsize $j$};
\node [centered, rotate=0.] at (3.,6.4) {\scriptsize $k$};
\node [centered, rotate=0.] at (3.4,7.) {\scriptsize $j$};
\node [centered, rotate=0.] at (4.2,6.7) {\scriptsize $q$};
\node [centered, rotate=0.] at (2.2,6.1) {\scriptsize $p^\ast$};
\end{tikzpicture}
}

\newcommand{\Tail}[1]{
\begin{tikzpicture}[baseline={([yshift=-.8ex]current bounding box.center)},scale=1.]
\draw [decorate,decoration={snake,segment length=.12cm, amplitude=.04cm}] (0, 0) -- (0, 1.5);
\node [right, rotate=0.] at (0.,.75) {\scriptsize ${#1}$};
\end{tikzpicture}
}

\newcommand{\HalfBraidingA}{
\begin{tikzpicture}[baseline={([yshift=-.8ex]current bounding box.center)},scale=1.]
\draw [->-] (0, 0) -- (0, .7);
\draw [->-] (.5, 0) -- (.5, .7);
\draw [] (-.2, .7) -- (.7, .7) -- (.7, 1.3) -- (-.2, 1.3) -- (-.2, .7);
\draw [->-] (0, 1.3) -- (0, 2.);
\draw [->-] (.5, 1.3) -- (.5, 2.);
\node [centered, rotate=0.] at (-.3, .2) {\scriptsize $X_J$};
\node [centered, rotate=0.] at (.7, .2) {\scriptsize $y$};
\node [centered, rotate=0.] at (-.2, 1.8) {\scriptsize $y$};
\node [centered, rotate=0.] at (.8, 1.8) {\scriptsize $X_J$};
\node [centered, rotate=0.] at (.25, 1.) {\scriptsize $c_{X_J, y}$};
\end{tikzpicture}\qquad =\qquad
\begin{tikzpicture}[baseline={([yshift=-.8ex]current bounding box.center)},scale=1.]
\draw [->-] (1, 0) -- (1, 1);
\draw [->-] (1, 1) -- (1, 2);
\draw [->-] (.5, 0) arc [start angle = 180, end angle = 100, x radius = .5, y radius = 1.];
\draw [-<-] (1.5, 2) arc [start angle = 0, end angle = -80, x radius = .5, y radius = 1.];
\node [left, rotate=0.] at (0.5, .25) {\scriptsize $X_J$};
\node [right, rotate=0.] at (1.5, 1.75) {\scriptsize $X_J$};
\node [centered, rotate=0.] at (1.2, .2) {\scriptsize $y$};
\node [centered, rotate=0.] at (.8, 1.8) {\scriptsize $y$};
\end{tikzpicture}
}

\newcommand{\HalfBraidingB}{
\begin{tikzpicture}[baseline={([yshift=-.8ex]current bounding box.center)},scale=.8]
\draw [->-] (1, 0) -- (1, 1);
\draw [->-] (1, 1) -- (1, 2);
\draw [->-] (.5, 0) arc [start angle = 180, end angle = 100, x radius = .5, y radius = 1.];
\draw [-<-] (1.5, 2) arc [start angle = 0, end angle = -80, x radius = .5, y radius = 1.];
\node [left, rotate=0.] at (0.5, .25) {\scriptsize $X_J$};
\node [right, rotate=0.] at (1.5, 1.75) {\scriptsize $X_J$};
\node [centered, rotate=0.] at (1.2, .2) {\scriptsize $y$};
\node [centered, rotate=0.] at (.8, 1.8) {\scriptsize $y$};
\end{tikzpicture}\ = \ \bigoplus_{p,q\in L_J}\ \bigoplus_{k\in L_\Fus}\ \sqrt{\frac{d_k}{d_y}}\ z_{pqy}^{J;k}\quad
\begin{tikzpicture}[baseline={([yshift=-.8ex]current bounding box.center)},scale=.8]
\draw [->-] (1, 0) -- (1, .8);
\draw [->-] (1, 1.2) -- (1, 2);
\draw [->-] (1, .8) -- (1, 1.2);
\draw [->-] (.5, 0) arc [start angle = 180, end angle = 90, x radius = .5, y radius = .7];
\draw [-<-] (1.5, 2) arc [start angle = 0, end angle = -90, x radius = .5, y radius = .7];
\node [left, rotate=0.] at (0.5, .25) {\scriptsize $p$};
\node [right, rotate=0.] at (1.5, 1.75) {\scriptsize $q$};
\node [centered, rotate=0.] at (1.2, .2) {\scriptsize $y$};
\node [centered, rotate=0.] at (.8, 1.8) {\scriptsize $y$};
\node [centered, rotate=0.] at (1.2, 1.) {\scriptsize $k$};
\end{tikzpicture}
}

\newcommand{\HalfBraidingC}{
\begin{tikzpicture}[baseline={([yshift=-.8ex]current bounding box.center)},scale=1]
\draw [->-] (1.15, .4) arc (-60:30:.5);
\draw [->-] (.6, .1) -- (.9, .3);
\draw [->-] (1.2, 1.3) -- (1., 1.6);

\draw [->-] (1, 0) -- (1, 1);
\draw [->-] (.2, 1.5) -- (1, 1);
\draw [->-] (1.8, 1.5) -- (1, 1);
\node [centered, rotate=0.] at (1.15, 0.) {\scriptsize $y$};
\node [centered, rotate=0.] at (.2, 1.3) {\scriptsize $u$};
\node [centered, rotate=0.] at (1.8, 1.3) {\scriptsize $v$};
\node [centered, rotate=0.] at (.4, 0.) {\scriptsize $p$};
\node [centered, rotate=0.] at (.8, 1.7) {\scriptsize $q$};
\end{tikzpicture}\qquad = \qquad 
\begin{tikzpicture}[baseline={([yshift=-.8ex]current bounding box.center)},scale=1]
\draw [->] (.5, .5) arc (-170:-240:1.);
\draw [->] (.5, .5) arc (-170:-190:1.);
\fill [white] (.55, 1.3) circle (.2);

\draw [->-] (1, 0) -- (1, 1);
\draw [->-] (.2, 1.5) -- (1, 1);
\draw [->-] (1.8, 1.5) -- (1, 1);
\node [centered, rotate=0.] at (1.15, 0.) {\scriptsize $y$};
\node [centered, rotate=0.] at (.2, 1.3) {\scriptsize $u$};
\node [centered, rotate=0.] at (1.8, 1.3) {\scriptsize $v$};
\node [centered, rotate=0.] at (.5, .3) {\scriptsize $p$};
\node [centered, rotate=0.] at (1.1, 1.6) {\scriptsize $q$};
\end{tikzpicture}
}

\newcommand{\HalfBraidingD}{
\begin{tikzpicture}[baseline={([yshift=-.8ex]current bounding box.center)},scale=.8]
\draw [->-] (1, 0) -- (1, 1);
\draw [->-] (1, 1) -- (1, 2);
\draw [->-] (.5, 0) arc [start angle = 180, end angle = 100, x radius = .5, y radius = 1.];
\draw [-<-] (1.5, 2) arc [start angle = 0, end angle = -80, x radius = .5, y radius = 1.];
\node [left, rotate=0.] at (0.5, .25) {\scriptsize $p$};
\node [right, rotate=0.] at (1.5, 1.75) {\scriptsize $q$};
\node [centered, rotate=0.] at (1.2, .2) {\scriptsize $j_E$};
\node [centered, rotate=0.] at (.8, 1.8) {\scriptsize $j_E$};
\end{tikzpicture} =
\sum_{k\in L_\Fus} \sqrt{\frac{d_k}{d_{j_E}}}\ {z_{pqj_E}^{J;k}} 
\begin{tikzpicture}[baseline={([yshift=-.8ex]current bounding box.center)},scale=.8]
\draw [->-] (1, 0) -- (1, .8);
\draw [->-] (1, 1.2) -- (1, 2);
\draw [->-] (1, .8) -- (1, 1.2);
\draw [->-] (.5, 0) arc [start angle = 180, end angle = 90, x radius = .5, y radius = .7];
\draw [-<-] (1.5, 2) arc [start angle = 0, end angle = -90, x radius = .5, y radius = .7];
\node [left, rotate=0.] at (0.5, .25) {\scriptsize $p$};
\node [right, rotate=0.] at (1.5, 1.75) {\scriptsize $q$};
\node [centered, rotate=0.] at (1.2, .2) {\scriptsize $j_E$};
\node [centered, rotate=0.] at (.8, 1.8) {\scriptsize $j_E$};
\node [centered, rotate=0.] at (1.2, 1.) {\scriptsize $k$};
\end{tikzpicture} =
W_E^{J;pq}\
\begin{tikzpicture}[baseline={([yshift=-.8ex]current bounding box.center)},scale=.8]
\draw [->-] (1, 0) -- (1, 2);
\node [centered, rotate=0.] at (1.3, 1.) {\scriptsize $j_E$};
\end{tikzpicture}\ .
}

\newcommand{\NonAbA}{
\begin{tikzpicture}[baseline={([yshift=-.8ex]current bounding box.center)},scale=.8]
\draw [->-] (3.2,5.6) -- (3.2,6.1);
\draw [->-] (3.2,6.1) -- (3.2,6.7);
\draw [->-] (3.2,6.7) -- (3.2,7.2);
\draw [->-] (3.2,7.2) -- (3.9,7.6);
\draw [->-] (2.5,7.6) -- (3.2,7.2);
\draw [->-] (2.5,5.2) -- (3.2,5.6);
\draw [->-] (3.2,5.6) -- (3.9,5.2);
\draw [decorate,decoration={snake,segment length=.12cm, amplitude=.04cm}, ->] (3.2,6.7) -- (4.,6.7);
\draw [decorate,decoration={snake,segment length=.12cm, amplitude=.04cm}, ->] (3.2,6.1) -- (2.4,6.1);
\node [centered, rotate=0.] at (2.4,5.) {\scriptsize $e_n$};
\node [centered, rotate=0.] at (4.1,5.) {\scriptsize $i_n$};
\node [centered, rotate=0.] at (4.1,7.8) {\scriptsize $i_2$};
\node [centered, rotate=0.] at (2.4,7.8) {\scriptsize $e_1$};
\node [centered, rotate=0.] at (3.,5.8) {\scriptsize $j$};
\node [centered, rotate=0.] at (3.,6.4) {\scriptsize $j$};
\node [centered, rotate=0.] at (3.4,7.) {\scriptsize $j$};
\node [centered, rotate=0.] at (4.4,6.7) {\scriptsize $1_{\mathcal{I}, 1}$};
\node [centered, rotate=0.] at (2.,6.1) {\scriptsize $1_{\mathcal{I},1}$};
\end{tikzpicture}
}

\newcommand{\NonAbB}{
\begin{tikzpicture}[baseline={([yshift=-.8ex]current bounding box.center)},scale=.8]
\draw [->-] (3.2,5.6) -- (3.2,6.1);
\draw [->-] (3.2,6.1) -- (3.2,6.7);
\draw [->-] (3.2,6.7) -- (3.2,7.2);
\draw [->-] (3.2,7.2) -- (3.9,7.6);
\draw [->-] (2.5,7.6) -- (3.2,7.2);
\draw [->-] (2.5,5.2) -- (3.2,5.6);
\draw [->-] (3.2,5.6) -- (3.9,5.2);
\draw [decorate,decoration={snake,segment length=.12cm, amplitude=.04cm}, ->] (3.2,6.7) -- (4.,6.7);
\draw [decorate,decoration={snake,segment length=.12cm, amplitude=.04cm}, ->] (3.2,6.1) -- (2.4,6.1);
\node [centered, rotate=0.] at (2.4,5.) {\scriptsize $e_n$};
\node [centered, rotate=0.] at (4.1,5.) {\scriptsize $i_n$};
\node [centered, rotate=0.] at (4.1,7.8) {\scriptsize $i_2$};
\node [centered, rotate=0.] at (2.4,7.8) {\scriptsize $e_1$};
\node [centered, rotate=0.] at (3.,5.8) {\scriptsize $j$};
\node [centered, rotate=0.] at (3.,6.4) {\scriptsize $k$};
\node [centered, rotate=0.] at (3.4,7.) {\scriptsize $j$};
\node [centered, rotate=0.] at (4.5,6.7) {\scriptsize $q_{J,\beta}$};
\node [centered, rotate=0.] at (1.9,6.) {\scriptsize $p^\ast_{J^\ast,\alpha}$};
\end{tikzpicture}
}

\begin{document}

\title{Anyon Condensation In Symmetry-Enriched Topological Phases: \textit{G}-Grading of Multifusion Categories}

\date{\today}
\author[a]{Nianrui Fu\footnote{Equal contribution}}
\author[a]{Siyuan Wang\footnote{Equal contribution}}
\author[a]{Yu Zhao\footnote{Corresponding author}}
\author[a,b]{Yidun Wan\footnote{Corresponding author}}
\affiliation[a]{State Key Laboratory of Surface Physics, Department of Physics, Center for Field Theory and Particle Physics, and Institute for Nanoelectronic devices and Quantum computing, Fudan University, Shanghai 200433, China}
\affiliation[b]{Hefei National Laboratory, Hefei 230088, China}
\emailAdd{nrfu25@m.fudan.edu.cn, siyuanwang18@fudan.edu.cn,  yuzhao20@fudan.edu.cn, ydwan@fudan.edu.cn}

\abstract{Although anyon condensation is a standard mechanism for relating topological orders, anyon condensation in symmetry-enriched topological (SET) phases is more intricate because the condensate must also be compatible with the global symmetry. We study symmetry-preserving anyon condensation in SET phases described by the enlarged Hu--Geer--Wu (HGW) string-net model with multifusion-category input data. We show that a $G$-preserving condensation is characterized by a compatible grading of the input multifusion category, and that this grading constructs the multifusion-category input of the child SET phase. To make this construction concrete, we consider the case where the relevant data come from a finite group extension $E$ of $G$ by $N$: an $E$-graded fusion category induces a $G$-SET input by passing to the quotient symmetry $G$, and the resulting input naturally carries a compatible $N$-grading that implements the further condensation inside the SET phase while preserving $G$. We illustrate the construction using three quantum-double examples: the trivial extension $\mathbb{Z}_2\times\mathbb{Z}_2$, the non-Abelian semidirect product $S_3$, and the nontrivial central extension $\mathbb{Z}_4$. The $\mathbb{Z}_4$ example further shows that symmetry fractionalized anyons do not obstruct symmetry-preserving condensation once the condensate is treated as a physical coherent state.}

\maketitle

\flushbottom
\section{Introduction}

Anyon condensation in $(2+1)$-dimensional topological orders has been studied from many complementary viewpoints, including condensable algebras, gapped boundaries, and exactly solvable lattice models \cite{Bais2002,Bais2003,Bais2009a,Kong2013,Barkeshli2010,Eli2010,Eliens2013,Burnell2011,Burnell2012,Burnell2018,Neupert2016,Hu2022,lin2023,zhao2025nonabelian}; it has also been realized in the Hu--Geer--Wu (HGW) string-net model, leading to a Landau-Ginzburg-Higgs paradigm of topological orders \cite{zhao2025c}. It is known that when spontaneous symmetry breaking is not triggered by a given anyon condensation in a parent topological order, the resulting child order is a symmetry-enriched topological (SET) order \cite{Hung2013,bischoff2019spontaneous}. While SET orders have been well understood, both formally \cite{yao2010symmetry,Mesaros2011,Lu2013,Hung2012a,Hung2013,Essin2012,SongHermele2014,Huang2013,Barkeshli2014c,Ning2018,cheng2017exactly} and in lattice models \cite{williamson2017a,lee2018,heinrich2016symmetry,fu2026symmetry}, anyon condensation in SET orders has rarely been addressed or understood. A natural question to ask is: Can we perform anyon condensation in a parent SET order while preserving the parent symmetry, and if so, what kind of condensation is allowed? Subsequent questions follow: What is the child order after such a symmetry-preserving anyon condensation? What is the most appropriate mathematical structure for describing both the condensation and the child order? These questions were partially answered in Ref.\cite{bischoff2019spontaneous} by means of group extensions in an abstract sense, similar to the conventional study of anyon condensation by means of condensable algebras in the modular tensor categories characterizing pure topological orders. Such abstract methods overlook important nuances of anyon condensation that can only be manifest in lattice models, where concrete physical Hilbert spaces are explicit. This is why the conclusion in Ref.\cite{bischoff2019spontaneous} that anyons carrying fractionalized symmetry charges in an SET order cannot condense without breaking the symmetry does not hold when we consider a lattice model realization of the SET order, where anyons are always created in pairs. These pairs are described by concrete Hilbert-space states that are invariant under the symmetry action. Since we have constructed lattice models of SET orders by taking multifusion categories as the input data, it is more physically appropriate to study anyon condensation in SET orders at the level of the input data of the multifusion Hu--Geer--Wu string-net model \cite{Hu2018,zhao2022,fu2026symmetry,fu2026nonlinear}. The key conclusions are:
\begin{itemize}
    \item An SET phase described by an enlarged HGW model with input multifusion category $\mathcal{M}$ admits a symmetry-preserving anyon condensation precisely when $\mathcal{M}$ admits a compatible group grading. This grading packages the condensate uniformly across all symmetry domains and domain walls.
    \item The child SET phase obtained from this condensation is again described by an enlarged HGW model. Its input multifusion category is the one constructed from the grading of $\mathcal{M}$, denoted schematically by $\mathcal{M}'={\M}^{N}_{\mathcal{M}}$ when the relevant grading group is $N$.
\end{itemize}
In this paper, we consider only the case in which the relevant global symmetries are described by finite groups. We formulate the compatibility condition required for $G$-graded multifusion categories to be well defined. Consequently, SET phases described by this class of categories support symmetry-preserving anyon condensation. We also present the explicit form of the input multifusion category for the resulting child SET phase after anyon condensation.

\section{Background}\label{sec:background}

A pure topological order may be realized by a string-net model whose input is a unitary fusion category $\mathscr{F}$; the anyon theory of the model is captured by the Drinfeld center $\mathcal{Z}(\mathscr{F})$ \cite{Wen1990a,Wen1991,Turaev1994,Drinfeld2010,Rowell2009,kong2014braided,Levin2004,Kitaev2006,Hu2018,kong2022invitationtopologicalorderscategory}. SET phases enrich this picture by coupling the intrinsic topological order to a global symmetry. Algebraically, SET data are often organized using $G$-crossed braided extensions, symmetry defects, and gauging procedures \cite{Mesaros2011,Lu2013,Hung2012a,Barkeshli2014c,Ning2018,cheng2017exactly,Cui2016a}. In lattice constructions, related SET models can be obtained by replacing the input fusion category with an appropriate multifusion category \cite{Chang2015,fu2026symmetry}.

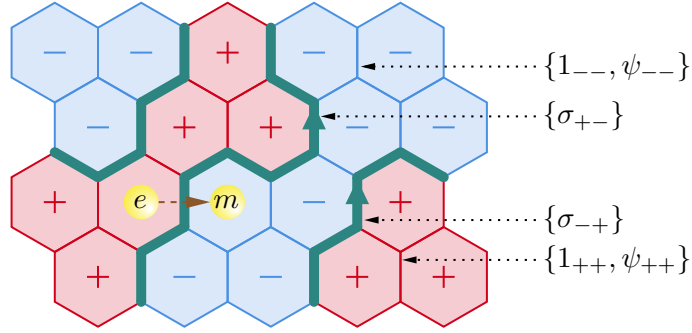
\begin{figure}[htbp]
    \centering
  
\tikzset {_5z4c4i2gc/.code = {\pgfsetadditionalshadetransform{ \pgftransformshift{\pgfpoint{-198 bp } { 158.4 bp }  }  \pgftransformscale{1.32 }  }}}
\pgfdeclareradialshading{_jom2r7lj4}{\pgfpoint{160bp}{-128bp}}{rgb(0bp)=(1,1,1);
rgb(0bp)=(1,1,1);
rgb(25bp)=(0.97,0.91,0.11);
rgb(400bp)=(0.97,0.91,0.11)}

  
\tikzset {_hdshpkk6j/.code = {\pgfsetadditionalshadetransform{ \pgftransformshift{\pgfpoint{-198 bp } { 158.4 bp }  }  \pgftransformscale{1.32 }  }}}
\pgfdeclareradialshading{_xi35rvywz}{\pgfpoint{160bp}{-128bp}}{rgb(0bp)=(1,1,1);
rgb(0bp)=(1,1,1);
rgb(25bp)=(0.97,0.91,0.11);
rgb(400bp)=(0.97,0.91,0.11)}
\tikzset{every picture/.style={line width=0.75pt}} 

\begin{tikzpicture}[x=0.75pt,y=0.75pt,yscale=-1,xscale=1]

\draw  [color={rgb, 255:red, 74; green, 144; blue, 226 }  ,draw opacity=1 ][fill={rgb, 255:red, 74; green, 144; blue, 226 }  ,fill opacity=0.2 ] (201.65,47.5) -- (180,60) -- (158.35,47.5) -- (158.35,22.5) -- (180,10) -- (201.65,22.5) -- cycle ;
\draw  [color={rgb, 255:red, 74; green, 144; blue, 226 }  ,draw opacity=1 ][fill={rgb, 255:red, 74; green, 144; blue, 226 }  ,fill opacity=0.2 ] (244.95,47.5) -- (223.3,60) -- (201.65,47.5) -- (201.65,22.5) -- (223.3,10) -- (244.95,22.5) -- cycle ;
\draw  [color={rgb, 255:red, 74; green, 144; blue, 226 }  ,draw opacity=1 ][fill={rgb, 255:red, 74; green, 144; blue, 226 }  ,fill opacity=0.2 ] (223.3,85) -- (201.65,97.5) -- (180,85) -- (180,60) -- (201.65,47.5) -- (223.3,60) -- cycle ;
\draw  [color={rgb, 255:red, 208; green, 2; blue, 27 }  ,draw opacity=1 ][fill={rgb, 255:red, 208; green, 2; blue, 27 }  ,fill opacity=0.2 ] (266.6,85) -- (244.95,97.5) -- (223.3,85) -- (223.3,60) -- (244.95,47.5) -- (266.6,60) -- cycle ;
\draw  [color={rgb, 255:red, 208; green, 2; blue, 27 }  ,draw opacity=1 ][fill={rgb, 255:red, 208; green, 2; blue, 27 }  ,fill opacity=0.2 ] (288.25,47.5) -- (266.6,60) -- (244.95,47.5) -- (244.95,22.5) -- (266.6,10) -- (288.25,22.5) -- cycle ;
\draw  [color={rgb, 255:red, 208; green, 2; blue, 27 }  ,draw opacity=1 ][fill={rgb, 255:red, 208; green, 2; blue, 27 }  ,fill opacity=0.2 ] (309.9,85) -- (288.25,97.5) -- (266.6,85) -- (266.6,60) -- (288.25,47.5) -- (309.9,60) -- cycle ;
\draw  [color={rgb, 255:red, 74; green, 144; blue, 226 }  ,draw opacity=1 ][fill={rgb, 255:red, 74; green, 144; blue, 226 }  ,fill opacity=0.2 ] (331.55,47.5) -- (309.9,60) -- (288.25,47.5) -- (288.25,22.5) -- (309.9,10) -- (331.55,22.5) -- cycle ;
\draw  [color={rgb, 255:red, 74; green, 144; blue, 226 }  ,draw opacity=1 ][fill={rgb, 255:red, 74; green, 144; blue, 226 }  ,fill opacity=0.2 ] (353.21,85) -- (331.55,97.5) -- (309.9,85) -- (309.9,60) -- (331.55,47.5) -- (353.21,60) -- cycle ;
\draw  [color={rgb, 255:red, 74; green, 144; blue, 226 }  ,draw opacity=1 ][fill={rgb, 255:red, 74; green, 144; blue, 226 }  ,fill opacity=0.2 ] (374.86,47.5) -- (353.21,60) -- (331.55,47.5) -- (331.55,22.5) -- (353.21,10) -- (374.86,22.5) -- cycle ;
\draw  [color={rgb, 255:red, 74; green, 144; blue, 226 }  ,draw opacity=1 ][fill={rgb, 255:red, 74; green, 144; blue, 226 }  ,fill opacity=0.2 ] (396.51,85) -- (374.86,97.5) -- (353.21,85) -- (353.21,60) -- (374.86,47.5) -- (396.51,60) -- cycle ;
\draw  [color={rgb, 255:red, 208; green, 2; blue, 27 }  ,draw opacity=1 ][fill={rgb, 255:red, 208; green, 2; blue, 27 }  ,fill opacity=0.2 ] (201.65,122.5) -- (180,135) -- (158.35,122.5) -- (158.35,97.5) -- (180,85) -- (201.65,97.5) -- cycle ;
\draw  [color={rgb, 255:red, 208; green, 2; blue, 27 }  ,draw opacity=1 ][fill={rgb, 255:red, 208; green, 2; blue, 27 }  ,fill opacity=0.2 ] (244.95,122.5) -- (223.3,135) -- (201.65,122.5) -- (201.65,97.5) -- (223.3,85) -- (244.95,97.5) -- cycle ;
\draw  [color={rgb, 255:red, 208; green, 2; blue, 27 }  ,draw opacity=1 ][fill={rgb, 255:red, 208; green, 2; blue, 27 }  ,fill opacity=0.2 ] (223.3,160) -- (201.65,172.5) -- (180,160) -- (180,135) -- (201.65,122.5) -- (223.3,135) -- cycle ;
\draw  [color={rgb, 255:red, 74; green, 144; blue, 226 }  ,draw opacity=1 ][fill={rgb, 255:red, 74; green, 144; blue, 226 }  ,fill opacity=0.2 ] (266.6,160) -- (244.95,172.5) -- (223.3,160) -- (223.3,135) -- (244.95,122.5) -- (266.6,135) -- cycle ;
\draw  [color={rgb, 255:red, 74; green, 144; blue, 226 }  ,draw opacity=1 ][fill={rgb, 255:red, 74; green, 144; blue, 226 }  ,fill opacity=0.2 ] (288.25,122.5) -- (266.6,135) -- (244.95,122.5) -- (244.95,97.5) -- (266.6,85) -- (288.25,97.5) -- cycle ;
\draw  [color={rgb, 255:red, 74; green, 144; blue, 226 }  ,draw opacity=1 ][fill={rgb, 255:red, 74; green, 144; blue, 226 }  ,fill opacity=0.2 ] (309.9,160) -- (288.25,172.5) -- (266.6,160) -- (266.6,135) -- (288.25,122.5) -- (309.9,135) -- cycle ;
\draw  [color={rgb, 255:red, 74; green, 144; blue, 226 }  ,draw opacity=1 ][fill={rgb, 255:red, 74; green, 144; blue, 226 }  ,fill opacity=0.2 ] (331.55,122.5) -- (309.9,135) -- (288.25,122.5) -- (288.25,97.5) -- (309.9,85) -- (331.55,97.5) -- cycle ;
\draw  [color={rgb, 255:red, 208; green, 2; blue, 27 }  ,draw opacity=1 ][fill={rgb, 255:red, 208; green, 2; blue, 27 }  ,fill opacity=0.2 ] (353.21,160) -- (331.55,172.5) -- (309.9,160) -- (309.9,135) -- (331.55,122.5) -- (353.21,135) -- cycle ;
\draw  [color={rgb, 255:red, 208; green, 2; blue, 27 }  ,draw opacity=1 ][fill={rgb, 255:red, 208; green, 2; blue, 27 }  ,fill opacity=0.2 ] (374.86,122.5) -- (353.21,135) -- (331.55,122.5) -- (331.55,97.5) -- (353.21,85) -- (374.86,97.5) -- cycle ;
\draw  [color={rgb, 255:red, 208; green, 2; blue, 27 }  ,draw opacity=1 ][fill={rgb, 255:red, 208; green, 2; blue, 27 }  ,fill opacity=0.2 ] (396.51,160) -- (374.86,172.5) -- (353.21,160) -- (353.21,135) -- (374.86,122.5) -- (396.51,135) -- cycle ;
\draw [color={rgb, 255:red, 74; green, 144; blue, 226 }  ,draw opacity=1 ][fill={rgb, 255:red, 74; green, 144; blue, 226 }  ,fill opacity=0.2 ]   (239.45,147.5) -- (250.45,147.5) ;
\draw [color={rgb, 255:red, 74; green, 144; blue, 226 }  ,draw opacity=1 ][fill={rgb, 255:red, 74; green, 144; blue, 226 }  ,fill opacity=0.2 ]   (261.1,110) -- (272.11,110) ;
\draw [color={rgb, 255:red, 74; green, 144; blue, 226 }  ,draw opacity=1 ][fill={rgb, 255:red, 74; green, 144; blue, 226 }  ,fill opacity=0.2 ]   (282.75,147.5) -- (293.76,147.5) ;
\draw [color={rgb, 255:red, 74; green, 144; blue, 226 }  ,draw opacity=1 ][fill={rgb, 255:red, 74; green, 144; blue, 226 }  ,fill opacity=0.2 ]   (347.7,35) -- (358.71,35) ;
\draw [color={rgb, 255:red, 74; green, 144; blue, 226 }  ,draw opacity=1 ][fill={rgb, 255:red, 74; green, 144; blue, 226 }  ,fill opacity=0.2 ]   (304.4,35) -- (315.41,35) ;
\draw [color={rgb, 255:red, 74; green, 144; blue, 226 }  ,draw opacity=1 ][fill={rgb, 255:red, 74; green, 144; blue, 226 }  ,fill opacity=0.2 ]   (326.05,72.5) -- (337.06,72.5) ;
\draw [color={rgb, 255:red, 74; green, 144; blue, 226 }  ,draw opacity=1 ][fill={rgb, 255:red, 74; green, 144; blue, 226 }  ,fill opacity=0.2 ]   (369.35,72.5) -- (380.36,72.5) ;
\draw [color={rgb, 255:red, 74; green, 144; blue, 226 }  ,draw opacity=1 ][fill={rgb, 255:red, 74; green, 144; blue, 226 }  ,fill opacity=0.2 ]   (217.8,35) -- (228.8,35) ;
\draw [color={rgb, 255:red, 74; green, 144; blue, 226 }  ,draw opacity=1 ][fill={rgb, 255:red, 74; green, 144; blue, 226 }  ,fill opacity=0.2 ]   (196.15,72.5) -- (207.15,72.5) ;
\draw [color={rgb, 255:red, 74; green, 144; blue, 226 }  ,draw opacity=1 ][fill={rgb, 255:red, 74; green, 144; blue, 226 }  ,fill opacity=0.2 ]   (174.5,35) -- (185.5,35) ;
\draw [color={rgb, 255:red, 74; green, 144; blue, 226 }  ,draw opacity=1 ][fill={rgb, 255:red, 74; green, 144; blue, 226 }  ,fill opacity=0.2 ]   (304.4,110) -- (315.41,110) ;
\draw [color={rgb, 255:red, 208; green, 2; blue, 27 }  ,draw opacity=1 ]   (223.3,115.5) -- (223.3,104.5) ;
\draw [color={rgb, 255:red, 208; green, 2; blue, 27 }  ,draw opacity=1 ]   (244.95,78) -- (244.95,67) ;
\draw [color={rgb, 255:red, 208; green, 2; blue, 27 }  ,draw opacity=1 ]   (180,115.5) -- (180,104.5) ;
\draw [color={rgb, 255:red, 208; green, 2; blue, 27 }  ,draw opacity=1 ]   (201.65,153) -- (201.65,142) ;
\draw [color={rgb, 255:red, 208; green, 2; blue, 27 }  ,draw opacity=1 ][fill={rgb, 255:red, 208; green, 2; blue, 27 }  ,fill opacity=0.2 ]   (266.6,40.5) -- (266.6,29.5) ;
\draw [color={rgb, 255:red, 208; green, 2; blue, 27 }  ,draw opacity=1 ]   (288.25,78) -- (288.25,67) ;
\draw [color={rgb, 255:red, 208; green, 2; blue, 27 }  ,draw opacity=1 ][fill={rgb, 255:red, 208; green, 2; blue, 27 }  ,fill opacity=0.2 ]   (353.21,115.5) -- (353.21,104.5) ;
\draw [color={rgb, 255:red, 208; green, 2; blue, 27 }  ,draw opacity=1 ]   (331.55,153) -- (331.55,142) ;
\draw [color={rgb, 255:red, 208; green, 2; blue, 27 }  ,draw opacity=1 ][fill={rgb, 255:red, 208; green, 2; blue, 27 }  ,fill opacity=0.2 ]   (374.86,153) -- (374.86,142) ;
\draw [color={rgb, 255:red, 208; green, 2; blue, 27 }  ,draw opacity=1 ][fill={rgb, 255:red, 208; green, 2; blue, 27 }  ,fill opacity=0.2 ]   (326.05,147.5) -- (337.06,147.5) ;
\draw [color={rgb, 255:red, 208; green, 2; blue, 27 }  ,draw opacity=1 ][fill={rgb, 255:red, 208; green, 2; blue, 27 }  ,fill opacity=0.2 ]   (369.35,147.5) -- (380.36,147.5) ;
\draw [color={rgb, 255:red, 208; green, 2; blue, 27 }  ,draw opacity=1 ][fill={rgb, 255:red, 208; green, 2; blue, 27 }  ,fill opacity=0.2 ]   (347.7,110) -- (358.71,110) ;
\draw [color={rgb, 255:red, 208; green, 2; blue, 27 }  ,draw opacity=1 ][fill={rgb, 255:red, 208; green, 2; blue, 27 }  ,fill opacity=0.2 ]   (261.1,35) -- (272.11,35) ;
\draw [color={rgb, 255:red, 208; green, 2; blue, 27 }  ,draw opacity=1 ]   (196.15,147.5) -- (207.15,147.5) ;
\draw [color={rgb, 255:red, 208; green, 2; blue, 27 }  ,draw opacity=1 ]   (217.8,110) -- (228.8,110) ;
\draw [color={rgb, 255:red, 208; green, 2; blue, 27 }  ,draw opacity=1 ][fill={rgb, 255:red, 208; green, 2; blue, 27 }  ,fill opacity=0.2 ]   (239.45,72.5) -- (250.45,72.5) ;
\draw [color={rgb, 255:red, 208; green, 2; blue, 27 }  ,draw opacity=1 ][fill={rgb, 255:red, 208; green, 2; blue, 27 }  ,fill opacity=0.2 ]   (282.75,72.5) -- (293.76,72.5) ;
\draw [color={rgb, 255:red, 208; green, 2; blue, 27 }  ,draw opacity=1 ]   (174.5,110) -- (185.5,110) ;
\draw [color={rgb, 255:red, 44; green, 133; blue, 128 }  ,draw opacity=1 ][line width=3.75]    (223.3,135) -- (223.3,160) ;
\draw [color={rgb, 255:red, 44; green, 133; blue, 128 }  ,draw opacity=1 ][line width=3.75]    (244.95,122.5) -- (223.3,135) ;
\draw  [draw opacity=0][fill={rgb, 255:red, 44; green, 133; blue, 128 }  ,fill opacity=1 ] (220.8,160) .. controls (220.8,158.62) and (221.92,157.49) .. (223.3,157.49) .. controls (224.69,157.49) and (225.81,158.62) .. (225.81,160) .. controls (225.81,161.38) and (224.69,162.51) .. (223.3,162.51) .. controls (221.92,162.51) and (220.8,161.38) .. (220.8,160) -- cycle ;
\draw  [draw opacity=0][fill={rgb, 255:red, 44; green, 133; blue, 128 }  ,fill opacity=1 ] (220.8,135) .. controls (220.8,133.62) and (221.92,132.49) .. (223.3,132.49) .. controls (224.69,132.49) and (225.81,133.62) .. (225.81,135) .. controls (225.81,136.38) and (224.69,137.51) .. (223.3,137.51) .. controls (221.92,137.51) and (220.8,136.38) .. (220.8,135) -- cycle ;
\draw [color={rgb, 255:red, 44; green, 133; blue, 128 }  ,draw opacity=1 ][line width=3.75]    (244.95,97.5) -- (244.95,122.5) ;
\draw  [draw opacity=0][fill={rgb, 255:red, 44; green, 133; blue, 128 }  ,fill opacity=1 ] (242.45,122.5) .. controls (242.45,121.12) and (243.57,119.99) .. (244.95,119.99) .. controls (246.34,119.99) and (247.46,121.12) .. (247.46,122.5) .. controls (247.46,123.88) and (246.34,125.01) .. (244.95,125.01) .. controls (243.57,125.01) and (242.45,123.88) .. (242.45,122.5) -- cycle ;
\draw [color={rgb, 255:red, 44; green, 133; blue, 128 }  ,draw opacity=1 ][line width=3.75]    (266.6,85) -- (244.95,97.5) ;
\draw [color={rgb, 255:red, 44; green, 133; blue, 128 }  ,draw opacity=1 ][line width=3.75]    (223.3,60) -- (223.3,85) ;
\draw [color={rgb, 255:red, 44; green, 133; blue, 128 }  ,draw opacity=1 ][line width=3.75]    (244.95,22.5) -- (244.95,47.5) ;
\draw [color={rgb, 255:red, 44; green, 133; blue, 128 }  ,draw opacity=1 ][line width=3.75]    (288.25,22.5) -- (288.25,47.5) ;
\draw [color={rgb, 255:red, 44; green, 133; blue, 128 }  ,draw opacity=1 ][line width=3.75]    (331.55,97.5) -- (331.55,122.5) ;
\draw [shift={(331.55,98.7)}, rotate = 90] [fill={rgb, 255:red, 44; green, 133; blue, 128 }  ,fill opacity=1 ][line width=0.08]  [draw opacity=0] (12.32,-5.92) -- (0,0) -- (12.32,5.92) -- cycle    ;
\draw [color={rgb, 255:red, 44; green, 133; blue, 128 }  ,draw opacity=1 ][line width=3.75]    (309.9,135) -- (309.9,160) ;
\draw [color={rgb, 255:red, 44; green, 133; blue, 128 }  ,draw opacity=1 ][line width=3.75]    (331.55,122.5) -- (309.9,135) ;
\draw [color={rgb, 255:red, 44; green, 133; blue, 128 }  ,draw opacity=1 ][line width=3.75]    (353.21,85) -- (331.55,97.5) ;
\draw [color={rgb, 255:red, 44; green, 133; blue, 128 }  ,draw opacity=1 ][line width=3.75]    (244.95,47.5) -- (223.3,60) ;
\draw [color={rgb, 255:red, 44; green, 133; blue, 128 }  ,draw opacity=1 ][line width=3.75]    (309.9,85) -- (288.25,97.5) ;
\draw [color={rgb, 255:red, 44; green, 133; blue, 128 }  ,draw opacity=1 ][line width=3.75]    (309.9,60) -- (309.9,85) ;
\draw [shift={(309.9,61.2)}, rotate = 90] [fill={rgb, 255:red, 44; green, 133; blue, 128 }  ,fill opacity=1 ][line width=0.08]  [draw opacity=0] (12.32,-5.92) -- (0,0) -- (12.32,5.92) -- cycle    ;
\draw [color={rgb, 255:red, 44; green, 133; blue, 128 }  ,draw opacity=1 ][line width=3.75]    (288.25,97.5) -- (266.6,85) ;
\draw [color={rgb, 255:red, 44; green, 133; blue, 128 }  ,draw opacity=1 ][line width=3.75]    (201.65,97.5) -- (180,85) ;
\draw [color={rgb, 255:red, 44; green, 133; blue, 128 }  ,draw opacity=1 ][line width=3.75]    (309.9,60) -- (288.25,47.5) ;
\draw [color={rgb, 255:red, 44; green, 133; blue, 128 }  ,draw opacity=1 ][line width=3.75]    (374.86,97.5) -- (353.21,85) ;
\draw [color={rgb, 255:red, 44; green, 133; blue, 128 }  ,draw opacity=1 ][line width=3.75]    (223.3,85) -- (201.65,97.5) ;
\draw  [draw opacity=0][fill={rgb, 255:red, 44; green, 133; blue, 128 }  ,fill opacity=1 ] (242.45,97.5) .. controls (242.45,96.12) and (243.57,94.99) .. (244.95,94.99) .. controls (246.34,94.99) and (247.46,96.12) .. (247.46,97.5) .. controls (247.46,98.88) and (246.34,100.01) .. (244.95,100.01) .. controls (243.57,100.01) and (242.45,98.88) .. (242.45,97.5) -- cycle ;
\draw  [draw opacity=0][fill={rgb, 255:red, 44; green, 133; blue, 128 }  ,fill opacity=1 ] (264.1,85) .. controls (264.1,83.62) and (265.22,82.49) .. (266.6,82.49) .. controls (267.99,82.49) and (269.11,83.62) .. (269.11,85) .. controls (269.11,86.38) and (267.99,87.51) .. (266.6,87.51) .. controls (265.22,87.51) and (264.1,86.38) .. (264.1,85) -- cycle ;
\draw  [draw opacity=0][fill={rgb, 255:red, 44; green, 133; blue, 128 }  ,fill opacity=1 ] (285.75,97.5) .. controls (285.75,96.12) and (286.87,94.99) .. (288.25,94.99) .. controls (289.64,94.99) and (290.76,96.12) .. (290.76,97.5) .. controls (290.76,98.88) and (289.64,100.01) .. (288.25,100.01) .. controls (286.87,100.01) and (285.75,98.88) .. (285.75,97.5) -- cycle ;
\draw  [draw opacity=0][fill={rgb, 255:red, 44; green, 133; blue, 128 }  ,fill opacity=1 ] (307.4,85) .. controls (307.4,83.62) and (308.52,82.49) .. (309.9,82.49) .. controls (311.29,82.49) and (312.41,83.62) .. (312.41,85) .. controls (312.41,86.38) and (311.29,87.51) .. (309.9,87.51) .. controls (308.52,87.51) and (307.4,86.38) .. (307.4,85) -- cycle ;
\draw  [draw opacity=0][fill={rgb, 255:red, 44; green, 133; blue, 128 }  ,fill opacity=1 ] (307.4,60) .. controls (307.4,58.62) and (308.52,57.49) .. (309.9,57.49) .. controls (311.29,57.49) and (312.41,58.62) .. (312.41,60) .. controls (312.41,61.38) and (311.29,62.51) .. (309.9,62.51) .. controls (308.52,62.51) and (307.4,61.38) .. (307.4,60) -- cycle ;
\draw  [draw opacity=0][fill={rgb, 255:red, 44; green, 133; blue, 128 }  ,fill opacity=1 ] (285.75,47.5) .. controls (285.75,46.12) and (286.87,44.99) .. (288.25,44.99) .. controls (289.64,44.99) and (290.76,46.12) .. (290.76,47.5) .. controls (290.76,48.88) and (289.64,50.01) .. (288.25,50.01) .. controls (286.87,50.01) and (285.75,48.88) .. (285.75,47.5) -- cycle ;
\draw  [draw opacity=0][fill={rgb, 255:red, 44; green, 133; blue, 128 }  ,fill opacity=1 ] (242.45,47.5) .. controls (242.45,46.12) and (243.57,44.99) .. (244.95,44.99) .. controls (246.34,44.99) and (247.46,46.12) .. (247.46,47.5) .. controls (247.46,48.88) and (246.34,50.01) .. (244.95,50.01) .. controls (243.57,50.01) and (242.45,48.88) .. (242.45,47.5) -- cycle ;
\draw  [draw opacity=0][fill={rgb, 255:red, 44; green, 133; blue, 128 }  ,fill opacity=1 ] (220.8,60) .. controls (220.8,58.62) and (221.92,57.49) .. (223.3,57.49) .. controls (224.69,57.49) and (225.81,58.62) .. (225.81,60) .. controls (225.81,61.38) and (224.69,62.51) .. (223.3,62.51) .. controls (221.92,62.51) and (220.8,61.38) .. (220.8,60) -- cycle ;
\draw  [draw opacity=0][fill={rgb, 255:red, 44; green, 133; blue, 128 }  ,fill opacity=1 ] (220.8,85) .. controls (220.8,83.62) and (221.92,82.49) .. (223.3,82.49) .. controls (224.69,82.49) and (225.81,83.62) .. (225.81,85) .. controls (225.81,86.38) and (224.69,87.51) .. (223.3,87.51) .. controls (221.92,87.51) and (220.8,86.38) .. (220.8,85) -- cycle ;
\draw  [draw opacity=0][fill={rgb, 255:red, 44; green, 133; blue, 128 }  ,fill opacity=1 ] (199.14,97.5) .. controls (199.14,96.12) and (200.27,94.99) .. (201.65,94.99) .. controls (203.03,94.99) and (204.16,96.12) .. (204.16,97.5) .. controls (204.16,98.88) and (203.03,100.01) .. (201.65,100.01) .. controls (200.27,100.01) and (199.14,98.88) .. (199.14,97.5) -- cycle ;
\draw  [draw opacity=0][fill={rgb, 255:red, 44; green, 133; blue, 128 }  ,fill opacity=1 ] (177.49,85) .. controls (177.49,83.62) and (178.62,82.49) .. (180,82.49) .. controls (181.38,82.49) and (182.51,83.62) .. (182.51,85) .. controls (182.51,86.38) and (181.38,87.51) .. (180,87.51) .. controls (178.62,87.51) and (177.49,86.38) .. (177.49,85) -- cycle ;
\draw  [draw opacity=0][fill={rgb, 255:red, 44; green, 133; blue, 128 }  ,fill opacity=1 ] (307.4,135) .. controls (307.4,133.62) and (308.52,132.49) .. (309.9,132.49) .. controls (311.29,132.49) and (312.41,133.62) .. (312.41,135) .. controls (312.41,136.38) and (311.29,137.51) .. (309.9,137.51) .. controls (308.52,137.51) and (307.4,136.38) .. (307.4,135) -- cycle ;
\draw  [draw opacity=0][fill={rgb, 255:red, 44; green, 133; blue, 128 }  ,fill opacity=1 ] (307.4,160) .. controls (307.4,158.62) and (308.52,157.49) .. (309.9,157.49) .. controls (311.29,157.49) and (312.41,158.62) .. (312.41,160) .. controls (312.41,161.38) and (311.29,162.51) .. (309.9,162.51) .. controls (308.52,162.51) and (307.4,161.38) .. (307.4,160) -- cycle ;
\draw  [draw opacity=0][fill={rgb, 255:red, 44; green, 133; blue, 128 }  ,fill opacity=1 ] (329.05,122.5) .. controls (329.05,121.12) and (330.17,119.99) .. (331.55,119.99) .. controls (332.94,119.99) and (334.06,121.12) .. (334.06,122.5) .. controls (334.06,123.88) and (332.94,125.01) .. (331.55,125.01) .. controls (330.17,125.01) and (329.05,123.88) .. (329.05,122.5) -- cycle ;
\draw  [draw opacity=0][fill={rgb, 255:red, 44; green, 133; blue, 128 }  ,fill opacity=1 ] (329.05,97.5) .. controls (329.05,96.12) and (330.17,94.99) .. (331.55,94.99) .. controls (332.94,94.99) and (334.06,96.12) .. (334.06,97.5) .. controls (334.06,98.88) and (332.94,100.01) .. (331.55,100.01) .. controls (330.17,100.01) and (329.05,98.88) .. (329.05,97.5) -- cycle ;
\draw  [draw opacity=0][fill={rgb, 255:red, 44; green, 133; blue, 128 }  ,fill opacity=1 ] (350.7,85) .. controls (350.7,83.62) and (351.82,82.49) .. (353.21,82.49) .. controls (354.59,82.49) and (355.71,83.62) .. (355.71,85) .. controls (355.71,86.38) and (354.59,87.51) .. (353.21,87.51) .. controls (351.82,87.51) and (350.7,86.38) .. (350.7,85) -- cycle ;
\draw  [draw opacity=0][fill={rgb, 255:red, 44; green, 133; blue, 128 }  ,fill opacity=1 ] (372.35,97.5) .. controls (372.35,96.12) and (373.47,94.99) .. (374.86,94.99) .. controls (376.24,94.99) and (377.36,96.12) .. (377.36,97.5) .. controls (377.36,98.88) and (376.24,100.01) .. (374.86,100.01) .. controls (373.47,100.01) and (372.35,98.88) .. (372.35,97.5) -- cycle ;
\draw  [draw opacity=0][fill={rgb, 255:red, 44; green, 133; blue, 128 }  ,fill opacity=1 ] (285.75,22.5) .. controls (285.75,21.12) and (286.87,19.99) .. (288.25,19.99) .. controls (289.64,19.99) and (290.76,21.12) .. (290.76,22.5) .. controls (290.76,23.88) and (289.64,25.01) .. (288.25,25.01) .. controls (286.87,25.01) and (285.75,23.88) .. (285.75,22.5) -- cycle ;
\draw  [draw opacity=0][fill={rgb, 255:red, 44; green, 133; blue, 128 }  ,fill opacity=1 ] (242.45,22.5) .. controls (242.45,21.12) and (243.57,19.99) .. (244.95,19.99) .. controls (246.34,19.99) and (247.46,21.12) .. (247.46,22.5) .. controls (247.46,23.88) and (246.34,25.01) .. (244.95,25.01) .. controls (243.57,25.01) and (242.45,23.88) .. (242.45,22.5) -- cycle ;
\draw  [dash pattern={on 0.84pt off 2.51pt}]  (421.05,42) -- (335.28,42) ;
\draw [shift={(333.28,42)}, rotate = 360] [fill={rgb, 255:red, 0; green, 0; blue, 0 }  ][line width=0.08]  [draw opacity=0] (7.2,-1.8) -- (0,0) -- (7.2,1.8) -- cycle    ;
\draw  [dash pattern={on 0.84pt off 2.51pt}]  (421.05,119) -- (337.05,119) ;
\draw [shift={(335.05,119)}, rotate = 360] [fill={rgb, 255:red, 0; green, 0; blue, 0 }  ][line width=0.08]  [draw opacity=0] (7.2,-1.8) -- (0,0) -- (7.2,1.8) -- cycle    ;
\draw  [dash pattern={on 0.84pt off 2.51pt}]  (421.05,139) -- (356.05,139) ;
\draw [shift={(354.05,139)}, rotate = 360] [fill={rgb, 255:red, 0; green, 0; blue, 0 }  ][line width=0.08]  [draw opacity=0] (7.2,-1.8) -- (0,0) -- (7.2,1.8) -- cycle    ;
\draw  [dash pattern={on 0.84pt off 2.51pt}]  (421.05,67) -- (315.28,67) ;
\draw [shift={(313.28,67)}, rotate = 360] [fill={rgb, 255:red, 0; green, 0; blue, 0 }  ][line width=0.08]  [draw opacity=0] (7.2,-1.8) -- (0,0) -- (7.2,1.8) -- cycle    ;
\draw  [draw opacity=0][shading=_jom2r7lj4,_5z4c4i2gc] (214.57,110) .. controls (214.57,105.18) and (218.48,101.27) .. (223.3,101.27) .. controls (228.12,101.27) and (232.03,105.18) .. (232.03,110) .. controls (232.03,114.82) and (228.12,118.73) .. (223.3,118.73) .. controls (218.48,118.73) and (214.57,114.82) .. (214.57,110) -- cycle ;
\draw  [draw opacity=0][shading=_xi35rvywz,_hdshpkk6j] (257.87,110) .. controls (257.87,105.18) and (261.78,101.27) .. (266.6,101.27) .. controls (271.43,101.27) and (275.34,105.18) .. (275.34,110) .. controls (275.34,114.82) and (271.43,118.73) .. (266.6,118.73) .. controls (261.78,118.73) and (257.87,114.82) .. (257.87,110) -- cycle ;
\draw [color={rgb, 255:red, 139; green, 87; blue, 42 }  ,draw opacity=1 ] [dash pattern={on 2.25pt off 2.25pt}]  (232.03,110) -- (255.87,110) ;
\draw [shift={(257.87,110)}, rotate = 180] [fill={rgb, 255:red, 139; green, 87; blue, 42 }  ,fill opacity=1 ][line width=0.08]  [draw opacity=0] (12,-3) -- (0,0) -- (12,3) -- cycle    ;

\draw (423,33.07) node [anchor=north west][inner sep=0.75pt]    {$\{1_{--} ,\psi _{--}\}$};
\draw (423,109.8) node [anchor=north west][inner sep=0.75pt]    {$\{\sigma _{-+}\}$};
\draw (423,130.13) node [anchor=north west][inner sep=0.75pt]    {$\{1_{++} ,\psi _{++}\}$};
\draw (423,58.07) node [anchor=north west][inner sep=0.75pt]    {$\{\sigma _{+-}\}$};
\draw (218,105) node [anchor=north west][inner sep=0.75pt]    {$e$};
\draw (258,105) node [anchor=north west][inner sep=0.75pt]    {$m$};
\end{tikzpicture}
    \caption{Input data for an SET lattice model. Plaquette domains are labeled the elements of symmetry group. The red and blue regions represent two local domain types, and the thick green links mark the domain walls between them. In the toric-code SET with EM-exchange symmetry, crossing such a wall sends an $e$ anyon to an $m$ anyon. The same domain labels determine the block of the input multifusion category from which each edge label is chosen: edges inside $+$ and $-$ domains carry objects in the diagonal blocks $\mathcal{M}_{++}$ and $\mathcal{M}_{--}$, while oriented domain-wall edges carry objects in the off-diagonal blocks $\mathcal{M}_{+-}$ and $\mathcal{M}_{-+}$.}
    \label{fig:set-input-domain-wall}
\end{figure}

The reason for using a multifusion category as SET input is physical. An SET model must keep track not only of the intrinsic topological degrees of freedom inside a symmetry domain, but also of the degrees of freedom living on domain walls that implement the global symmetry action \cite{Bais2009,Kitaev2012a,Fuchs2013,Lan2014,Teo2013,Bombin2010,Buerschaper2013,aasen2020topological}. Consider the EM-exchange-symmetry-enriched $\mathbb{Z}_2$ toric code as an example. Seen in Fig. \ref{fig:set-input-domain-wall}, each plaquette is first assigned a symmetry-domain label, say $+$ or $-$. An edge between an $i$-plaquette and a $j$-plaquette is then labeled by a degree of freedom appropriate to that local environment. Thus, an edge inside a $+$ ($-$) domain carries a $++$ ($--$) label, and an edge lying on a domain wall carries a $+-$ or $-+$ label, depending on the orientation of the wall (Fig. \ref{fig:set-input-domain-wall}). The diagonal degrees of freedom describe toric-code input data inside a fixed domain, while the off-diagonal wall degrees of freedom carry the defects that exchange $e$ and $m$, as illustrated in Fig. \ref{fig:set-input-domain-wall}. These four types of local data naturally organize into the blocks of a multifusion category: an edge between an $i$-plaquette and a $j$-plaquette is labeled by an object of $\mathcal{M}_{ij}$, where $i,j\in\{+,-\}$. More generally, a fusion category with a simple tensor unit describes one uniform domain.

\begin{figure}[t]
    \centering
    \includegraphics[width=0.95\textwidth]{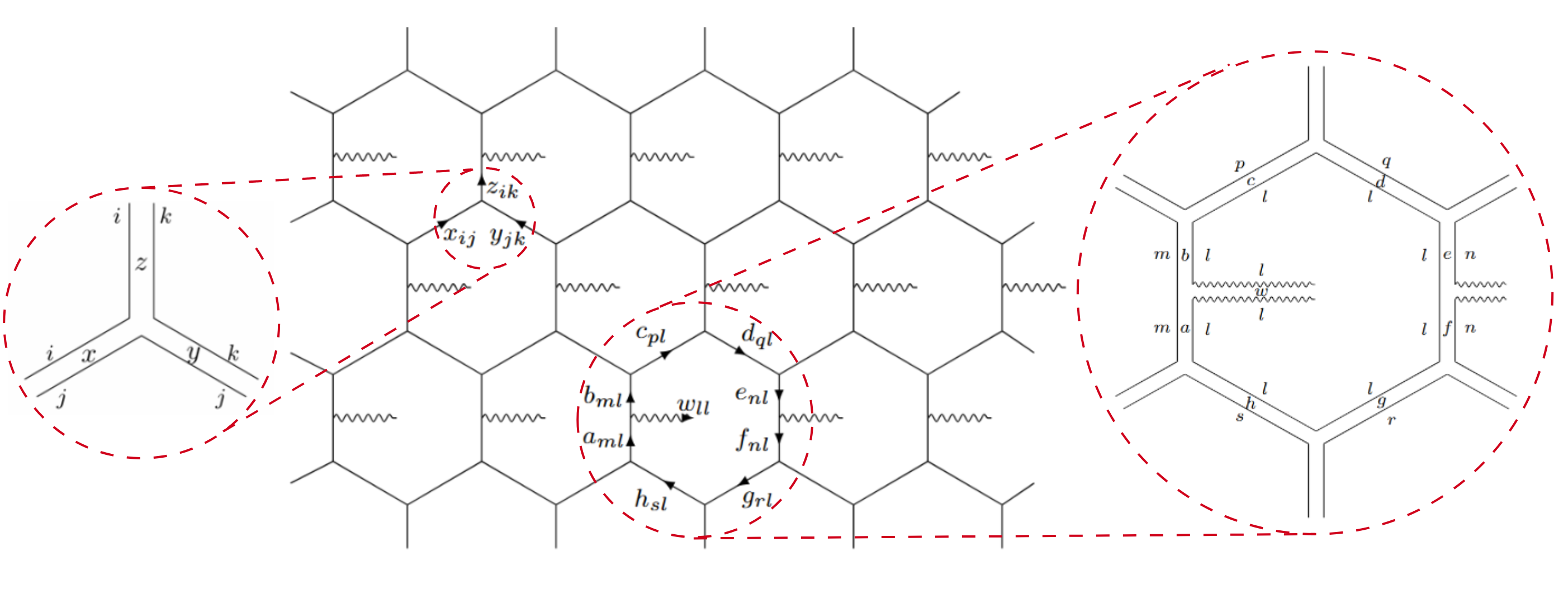}
    \caption{The fattened honeycomb lattice. Here, $x_{ij}\in\mathcal{M}_{ij}$, $y_{jk}\in\mathcal{M}_{jk}$, $z_{ik}\in\mathcal{M}_{ik}$, etc. The fusion rule $\mathcal{M}_{ij}\otimes\mathcal{M}_{kl}\subset\delta_{jk}\mathcal{M}_{il}$ allows each vertex to be fattened, as shown in the left figure. Multifusion indices are assigned to the doubled lines, with connected doubled lines carrying the same index. Consequently, after fattening the entire lattice, each plaquette is naturally labeled by a multifusion index, as illustrated in the right figure.}
    \label{fig:fatlattice}
\end{figure}

In general, the multifusion formulation is especially useful because it turns this rule of labeling the degrees of freedom across domains into algebra. A multifusion category has a decomposable tensor unit and a matrix-like decomposition
\begin{equation}
    \mathcal{M}=\bigoplus_{i,j}\mathcal{M}_{ij}.
\end{equation}
In the SET lattice model, the label $i$ is interpreted as spatial domain. A diagonal block $\mathcal{M}_{ii}$ describes degrees of freedom inside the $i$-domain, while an off-diagonal block $\mathcal{M}_{ij}$ describes the domain wall from sector $i$ to sector $j$. The fusion rule $\mathcal{M}_{ij}\otimes\mathcal{M}_{kl}\subset\delta_{jk}\mathcal{M}_{il}$ naturally motivates representing each vertex in the lattice model by a doubled-line structure, with the multifusion indices assigned to the doubled lines, as illustrated in Fig. \ref{fig:fatlattice}. In the doubled-line representation, we can see that each plaquette is labeled by a multifusion index (as depicted by Fig. \ref{fig:fatlattice}), which indicates the domain \cite{Chang2015,fu2026symmetry}.

This multifusion description can be viewed, in physical terms, as arising from a $G$-graded fusion category: each multifusion category that is a legal input data for an SET string-net model corresponds to such a grading. Any SET phase can be regarded as the child phase of an anyon condensation process. More precisely, every SET phase is associated with a unique parent pure topological order. Condensing a suitable set of anyons in the parent phase drives a phase transition to the corresponding child SET phase. At the input level, this condensation process is described by $G$-grading\footnote{In this paper, we only consider the case when anyon condensation yields group global symmetry.}. Specifically, if the input fusion category $\mathscr{F}$ of the parent phase admits anyon condensation, it must possess a $G$-graded structure
\begin{equation}\label{eq:grade}
    \mathscr{F} = \bigoplus_{g\in G}\mathscr{F}_g,
\end{equation}
where 
\begin{equation}\label{eq:gradecondition}
    \mathscr{F}_g\times\mathscr{F}_h\subset\mathscr{F}_{gh}.
\end{equation} 
The $G$-grading then determines the input multifusion category $\mathcal{M}$ of the child SET phase through
\begin{equation}\label{eq:setconstruction}
    \mathcal{M}_{g_i,g_j}=\mathscr{F}_{g_i^{-1}g_j}.
\end{equation}
The global symmetry group of the child SET phase is $G$. Appendix \ref{app:g-grading} details the definition of $G$-grading.

From the construction in Eq.~\eqref{eq:setconstruction}, the diagonal blocks with $g=h$ satisfy $\mathcal{M}_{g,g} \simeq \mathscr{F}_e$, where $e$ is the group identity. Consequently, the total tensor unit of $\mathcal{M}_\mathscr{F}^G$ consists of $|G|$ distinct copies of the original unit $\mathbb{1}^G \in \mathscr{F}_e$:
\begin{equation}
    \mathbb{1}_{\mathcal{M}} = \bigoplus_{g\in G}\mathbb{1}_{g,g},
\end{equation}
where $\mathbb{1}_{g,g}$ acts as the identity of the block $\mathcal{M}_{g,g}$. The non-neutral components $\mathscr{F}_{a\ne e}$ therefore do not appear as ordinary bulk degrees of freedom inside a domain. Instead, they occupy off-diagonal blocks and are confined to domain walls. In the HGW lattice picture, this explains why these confined degrees of freedom may decorate wall edges but cannot appear as isolated local bulk excitations in a uniform domain.

A basic example is the Ising category. Its simple objects are $\{1,\sigma,\psi\}$, with fusion rule $\sigma\otimes\sigma=1\oplus\psi$. It admits a $\mathbb{Z}_2$-grading, with $G=\{+,-\}$, given by
\begin{equation}\label{eq:ising grading}
    \mathscr{F}_+ = \{1,\psi\}, \quad \mathscr{F}_{-} = \{\sigma\}.
\end{equation}
At the level of topological orders, the doubled Ising theory admits a condensation of the boson $\psi\bar\psi$ associated with this grading. After this condensation, the neutral input is
\begin{equation}
    \mathscr{F}_+= \Vec(\mathbb{Z}_2),
\end{equation}
whose Drinfeld center is the $\mathbb{Z}_2$ Toric Code. The nontrivial component $\{\sigma\}$ is then reinterpreted as the $\mathbb{Z}_2$ symmetry-defect sector of the resulting SET. The corresponding multifusion category $\mathcal{M}_{\mathrm{Ising}}^{\mathbb{Z}_2}$ has diagonal blocks
\begin{equation}
    \mathcal{M}_{++}
    =
    \mathcal{M}_{--}
    =
    \{1,\psi\},
\end{equation}
Thus, each domain carries the input $\Vec(\mathbb{Z}_2)$ for the Toric Code. The off-diagonal blocks are
\begin{equation}
    \mathcal{M}_{+-}=\{\sigma_{+-}\},
    \qquad
    \mathcal{M}_{-+}=\{\sigma_{-+}\}.
\end{equation}
These off-diagonal blocks are one-dimensional symmetry domain walls separating the two $\mathbb{Z}_2$ domains. The multifusion category reads
\begin{equation}\label{eq:isng multifusion}
    \mathcal{M}_\text{Ising}^{\mathbb{Z}_2} = \begin{pmatrix}
        \{1_{++},\psi_{++}\} & \{\sigma_{+-}\} \\
        \{\sigma_{-+}\} & \{1_{--},\psi_{--}\}
    \end{pmatrix}.
\end{equation}
Mathematically, $\mathcal{M}_\mathscr{F}^G$ and $\mathscr{F}$ are Morita equivalent, so their string-net models have equivalent bulk centers, $\mathcal{Z}(\mathcal{M}_\mathscr{F}^G)\simeq\mathcal{Z}(\mathscr{F})$. Physically, however, they organize the same bulk theory in different ways. In the multifusion description, $\{1_{++},\psi_{++}\}$ and $\{1_{--},\psi_{--}\}$ describe two symmetry-breaking child domains, each locally governed by $\Vec(\mathbb{Z}_2)$ and hence supporting Toric Code bulk order. The off-diagonal $\{\sigma_{+-}\}$ and $\{\sigma_{-+}\}$ blocks describe the domain walls between these domains. The $\sigma$ object is the wall degree of freedom that implements the nontrivial $\mathbb{Z}_2$ symmetry action. Crossing the wall implements electromagnetic duality \cite{Hu2020,Wang2020}: an electric charge $e$ is transported to a magnetic flux $m$, and vice versa \cite{heinrich2016symmetry,fu2026symmetry}. Thus the multifusion category $\mathcal{M}_\text{Ising}^{\mathbb{Z}_2}$ \eqref{eq:isng multifusion} should be read as the input of a $\mathbb{Z}_2$-enriched Toric Code whose symmetry exchanges $e$ and $m$.

\section{Symmetry-preserving condensation from graded multifusion categories}\label{sec:setcondensation}

Now let's consider anyon condensation in SET phases. In an SET phase with global symmetry $G$ (called it a $G$-SET), any condensation that preserves $G$ must use a $G$-invariant condensate. Otherwise, the condensation selects a preferred symmetry sector and breaks the symmetry $G$. Therefore, such a condensation must lead to another global symmetry $N$ that is not part of $G$. This amounts to further grading (by $N$) the input multifusion category $\mathcal{M}_\mathscr{F}^G$ of the $G$-SET phase. If such a grading does not exist, the $G$-SET admits no further anyon condensation that preserves $G$.

\subsection{The general picture}

To facilitate the general formulation of the story above, let's continue to consider the previous example: the EM-exchange-symmetry-enriched $\Z_2$ toric code. Its input multifusion category $\mathcal{M}^{\mathbb{Z}_2}_{\text{Ising}}$ in \eqref{eq:isng multifusion} comes from the $\mathbb{Z}_2$-grading \eqref{eq:ising grading} of the Ising category, as reviewed in Section \ref{sec:background}. One might think that this SET has two possible condensates: $1+e$ and $1+m$, which are mutually exclusive because $e$ braids nontrivially with $m$. Nevertheless, the EM-exchange symmetry permits neither of these two would-be condensates, which inevitably breaks the EM-exchange symmetry. Thus, this SET phase admits no symmetry-preserving condensation. This physics is directly translated into the mathematics that the input multifusion category $\mathcal{M}^{\mathbb{Z}_2}_{\text{Ising}}$ bears no nontrivial grading\footnote{In general, every multifusion category $\M^G_\mathscr{F}$ has a \textbf{trivial} $G$-grading structure $[\M^G_\mathscr{F}]^{(g)}=\bigoplus_{g'\in G}(\M^G_\mathscr{F})_{g',g'g},\ g\in G$, which only encodes the existing global symmetry $G$. For anyon condensation, by contrast, we expect a new symmetry $N$ after condensation; hence what we need is a \textbf{nontrivial} grading of $\M^G_\mathscr{F}$ that grades the neutral component $\mathscr{F}_e$.}: If one were to grade $\mathcal{M}^{\mathbb{Z}_2}_{\text{Ising}}$, one would have to grade its neutral component  $\mathscr{F}_+= \Vec(\mathbb{Z}_2)$, such that
\begin{equation}\label{eq:z2grading}
    \Vec(\mathbb{Z}_2)_+=\{1\},\quad\Vec(\mathbb{Z}_2)_-=\{\psi\}.
\end{equation}
But this is impossible because the fusion rules $\sigma_{+-}\otimes\sigma_{-+}=1_{++}\oplus\psi_{++}$ ($\sigma_{-+}\otimes\sigma_{+-}=1_{--}\oplus\psi_{--}$) demand that $1_{++}$ ($1_{--}$) and $\psi_{++}$ ($\psi_{--}$) must lie in the same graded component.  

In general, for a $G$-SET phase admitting a $G$-preserving anyon condensation, its model's input multifusion category $\M^G_{\Fus}$ must bear a nontrivial $N$-grading:
\begin{equation}
    \M^G_{\Fus}=\bigoplus_{h\in N}[\M^G_{\Fus}]^{(h)},
\end{equation}
such that
\begin{equation}
    [\M^G_{\Fus}]^{(h)}\times [\M^G_{\Fus}]^{(\ell)}
    \subset [\M^G_{\Fus}]^{(h\ell)},\ \forall h,\ell\in N,
\end{equation}
where $N$ is a group. Each multifusion block of $\M^G_{\Fus}$ is divided into $|N|$ components under grading.

Given such a grading, we repeat the construction used in dealing with anyon condensation in fusion categories in \eqref{eq:setconstruction}. Define a new multifusion category $\M^{N}_{\M^G_{\Fus}}$ by
\begin{equation}
    \M^{N}_{\M^G_{\Fus}}
    :=
    \bigoplus_{h,\ell\in N}
    \left(\M^{N}_{\M^G_{\Fus}}\right)_{h,\ell},
\end{equation}
where the multifusion block components $(\M^{N}_{\M^G_{\Fus}})_{h,\ell}$ are defined by
\begin{equation}
    \left(\M^{N}_{\M^G_{\Fus}}\right)_{h,\ell}
    :=
    [\M^G_{\Fus}]^{(h^{-1}\ell)}.
\end{equation}
For $X_{h,\ell}\in (\M^{N}_{\M^G_{\Fus}})_{h,\ell}$ and $Y_{\ell,m}\in (\M^{N}_{\M^G_{\Fus}})_{\ell,m}$, we have
\begin{equation}
    X_{h,\ell}\otimes Y_{\ell,m}
    =
    (X\otimes Y)_{h,m}
    \in
    \left(\M^{N}_{\M^G_{\Fus}}\right)_{h,m},
\end{equation}
which is guaranteed by the grading condition. Physically, the new pair of indices $(h,\ell)$ refines the domain labels after condensation. Thus, an $N$-grading of $\M^G_{\Fus}$ is the input-level criterion for a symmetry-preserving condensation in the SET model with input $\M^G_{\Fus}$.

In what follows, for a parent $G$-SET phase that allows $G$-preserving anyon condensation, we shall construct the input multifusion category $\M^{N}_{\M^G_{\Fus}}$ of the child SET phase that acquires a further global symmetry $N$, as well as inheriting the $G$ symmetry from the parent. The total global symmetry turns out to be a group extension of $G$ by $N$, yielding a larger group $E$. Physically, this extension can be understood as a $2$-step condensation: The first anyon condensation occurs in a pure topological order with input fusion category $\Fus$, yielding the $G$-SET phase. The second condensation takes place in this $G$-SET phase without breaking $G$, enlarging the symmetry by $N$ and yielding an $E$-SET phase.  

\subsection{The 2-step condensation}

In general, the extended group $E$ from $G$ by $N$ takes the form as a twisted semidirect product
\begin{equation}\label{eq:groupextension}
    E=N\rtimes_{\varphi,\omega}G,
\end{equation}
with $\varphi:G\to\operatorname{Aut}(N)$, and $\omega\in H^2[G,N]$ being a 2-cocycle. It is worth noting that $\omega$ exists only if $N$ is Abelian.

Mathematically, the two-step condensation raised in the end of previous subsection can be organized as follows. Let $\phi:E\to G$ be the quotient map, with kernel $N$. Let $\mathscr{F}$ be any fusion category bearing an $E$-grading
\begin{equation}
    \mathscr{F}=\bigoplus_{a\in E}\mathscr{F}_a.
\end{equation}
The quotient map induces a $G$-grading
\begin{equation}
    \Fus=\bigoplus_{g\in G}\left(\bigoplus_{a\in \phi^{-1}(g)}\mathscr{F}_a\right),
\end{equation}
which encodes the first-step condensation. In physical terms, the first condensation is the condensation of the bosonic charge algebra associated with $\operatorname{Rep}(G)\subset \operatorname{Rep}(E)$, pulled back along $\phi$. It confines the $E$-fluxes outside $N$ and produces the $G$-SET phase -- SET phase with global symmetry $G$ -- in which we want to do the second condensation that preserves the $G$ symmetry. The input of this phase is $\mathcal{M}^{G}_{\mathscr{F}}$, which itself carries a nontrivial $N$-grading, so we can repeat the multifusion category construction and obtain
\begin{equation}\label{eq:two-step-result}
    \mathcal{M}^{N}_{\mathcal{M}^{G}_{\mathscr{F}}}
    \simeq
    \mathcal{M}^{E}_{\mathscr{F}}.
\end{equation}
Thus, the two-stage process yields the final input SET data, which agrees with the one obtained by grading directly with the full extension group. 

We now elaborate this general construction in two representative cases. The first is the split case, while the second includes is the twisted case.

\subsubsection{Case I: the split case}

The first case is the split extension, in which
\begin{equation}
    E=N\rtimes_{\varphi}G,
\end{equation}
where $\varphi:G\to \operatorname{Aut}(N)$ and
\begin{equation}
    (h,g)(h',g')=(h\varphi_g(h'),gg').
\end{equation}
Let $\mathscr{F}=\bigoplus_{(h,g)\in N\rtimes_\varphi G}\mathscr{F}_{(h,g)}$ be any $E$-graded fusion category. The quotient $G$-grading is
\begin{equation}
    \Fus=\bigoplus_{g\in G}\left(\bigoplus_{h\in N}\mathscr{F}_{(h,g)}\right).
\end{equation}
The first multifusion category is therefore
\begin{equation}\label{eq:first-M-semidirect}
    \left(\mathcal{M}^{G}_{\mathscr{F}}\right)_{g_1,g_2}
    =
    \left[\bigoplus_{h\in N}\left(\mathscr{F}_{(h,g_1^{-1}g_2)}\right)\right]_{g_1,g_2}.
\end{equation}
It admits the following nontrivial $N$-grading:
\begin{equation}\label{eq:semi-grading}
    \mathcal{M}^{(h_i)}
    =
    \bigoplus_{q,g\in G}
    \left(\mathscr{F}_{(\varphi_{q^{-1}}(h_i),g)}\right)_{q,qg}.
\end{equation}
To check the grading condition, take
\begin{equation}
    X\in\left(\mathscr{F}_{(\varphi_{q^{-1}}(h_i),g)}\right)_{q,qg},
    \qquad
    Y\in\left(\mathscr{F}_{(\varphi_{(qg)^{-1}}(h_j),g')}\right)_{qg,qgg'},
\end{equation}
their tensor product is nonzero only if the adjacent $G$-indices match. Since $\mathscr{F}$ is $E$-graded,
\begin{align}
    X\otimes Y
    &\in
    \left(\mathscr{F}_{(\varphi_{q^{-1}}(h_i),g)(\varphi_{(qg)^{-1}}(h_j),g')}\right)_{q,qgg'} \notag\\
    &=
    \left(\mathscr{F}_{(\varphi_{q^{-1}}(h_i)\varphi_g(\varphi_{g^{-1}q^{-1}}(h_j)),gg')}\right)_{q,qgg'} \notag\\
    &=
    \left(\mathscr{F}_{(\varphi_{q^{-1}}(h_ih_j),gg')}\right)_{q,qgg'}
    \subset \mathcal{M}^{(h_ih_j)}.
\end{align}
Hence, \eqref{eq:semi-grading} defines an $N$-grading.

The corresponding second multifusion category has blocks
\begin{align}
    \left(
    \mathcal{M}^{N}_{\mathcal{M}^{G}_{\mathscr{F}}}
    \right)_{h_1,h_2}
    &=
    \left[\bigoplus_{q,g\in G}
    \left(\mathscr{F}_{(\varphi_{q^{-1}}(h_1^{-1}h_2),g)}\right)_{q,qg}\right]_{h_1,h_2}.
\end{align}
Combining the two kinds of indices as $(h,q)\in N\rtimes_\varphi G$, this may be rewritten as
\begin{equation}
    \mathcal{M}^{N}_{\mathcal{M}^{G}_{\mathscr{F}}}\simeq\bigoplus_{h_1,h_2\in N,\ q,g\in G}
    \left(\mathscr{F}_{(\varphi_{q^{-1}}(h_1^{-1}h_2),g)}\right)_{(h_1,q),(h_2,qg)}.
\end{equation}
Since
\begin{equation}
    (h_1,q)(\varphi_{q^{-1}}(h_1^{-1}h_2),g)
    =
    (h_2,qg),
\end{equation}
We have
\begin{equation}
    \mathcal{M}^{N}_{\mathcal{M}^{G}_{\mathscr{F}}}\simeq\mathcal{M}^{E}_{\mathscr{F}}=\bigoplus_{h_1,h_2\in N,\ q,g\in G}
    \left(\mathscr{F}_{(\varphi_{q^{-1}}(h_1^{-1}h_2),g)}\right)_{(h_1,q),(h_2,qg)}.
\end{equation}
This proves \eqref{eq:two-step-result} for any $E$-graded fusion category $\mathscr{F}$ in the split semidirect product case.

\subsubsection{Case II: the twisted case}

The second case is the cocycle-twisted extension, where a nontrivial 2-cocycle $\omega\in H^2[G,N]$ enters the group law; here $N$ has to be Abelian:
\begin{equation}
    E=N\rtimes_{\varphi,\omega}G,
\end{equation}
with multiplication
\begin{equation}
    (h,g)(h',g')
    =
    (h+\varphi_g(h')+\omega(g,g'),gg').
\end{equation}
The cocycle condition is
\begin{equation}
    \omega(g_1,g_2)+\omega(g_1g_2,g_3)
    =
    \varphi_{g_1}(\omega(g_2,g_3))+\omega(g_1,g_2g_3).
\end{equation}
The quotient $G$-grading and the first multifusion category have the same form as in \eqref{eq:first-M-semidirect}. The nontrivial $N$-grading, however, must be shifted by the cocycle:
\begin{equation}\label{eq:twisted-grading}
    \mathcal{M}^{(h_i)}
    =
    \bigoplus_{q,g\in G}
    \left(\mathscr{F}_{(\varphi_{q^{-1}}(h_i-\omega(q,g)),g)}\right)_{q,qg}.
\end{equation}
Indeed, let
\begin{equation}
    X\in\left(\mathscr{F}_{(\varphi_{q^{-1}}(h_i-\omega(q,g)),g)}\right)_{q,qg},
\end{equation}
and
\begin{equation}
    Y\in\left(\mathscr{F}_{(\varphi_{(qg)^{-1}}(h_j-\omega(qg,g')),g')}\right)_{qg,qgg'}.
\end{equation}
Using the multiplication in $E$ and the cocycle condition gives
\begin{align}
    X\otimes Y
    &\in
    \left(\mathscr{F}_{(\varphi_{q^{-1}}(h_i-\omega(q,g)),g)
    (\varphi_{(qg)^{-1}}(h_j-\omega(qg,g')),g')}\right)_{q,qgg'} \notag\\
    &=
    \left(\mathscr{F}_{(\varphi_{q^{-1}}(h_i+h_j-\omega(q,gg')),gg')}\right)_{q,qgg'}
    \subset \mathcal{M}^{(h_i+h_j)}.
\end{align}
Thus, \eqref{eq:twisted-grading} defines an $N$-grading.

The corresponding second multifusion category has blocks
\begin{align}
    \left(
    \mathcal{M}^{N}_{\mathcal{M}^{G}_{\mathscr{F}}}
    \right)_{h_1,h_2}
    &=
    \left[\bigoplus_{q,g\in G}
    \left(\mathscr{F}_{(\varphi_{q^{-1}}(h_1^{-1}h_2-\omega(q,g)),g)}\right)_{q,qg}\right]_{h_1,h_2}.
\end{align}
Similar to case I, combining the two kinds of indices as $(h,q)\in N\rtimes_{\varphi,\omega}G$, the second multifusion category may be rewritten as
\begin{equation}
    \mathcal{M}^{N}_{\mathcal{M}^{G}_{\mathscr{F}}}\simeq\bigoplus_{h_1,h_2\in N,\ q,g\in G}
    \left(\mathscr{F}_{(\varphi_{q^{-1}}(h_2-h_1-\omega(q,g)),g)}\right)_{(h_1,q),(h_2,qg)}.
\end{equation}
Since
\begin{equation}
    (h_1,q)(\varphi_{q^{-1}}(h_2-h_1-\omega(q,g)),g)
    =
    (h_2,qg),
\end{equation}
we can see that $\mathcal{M}^{N}_{\mathcal{M}^{G}_{\mathscr{F}}}$ is again exactly $\mathcal{M}^{E}_{\mathscr{F}}$. In the special case where $\varphi$ is trivial, this describes a central extension.

\subsection{Three quantum-double examples}\label{sec:examples}

We now spell out the construction in three group-theoretic examples. In each case, the parent input is $\operatorname{Vec}(E)$, where $E$ is the group extension of $G$ by $N$, so the parent anyon theory is the Drinfeld center $\mathcal{Z}(\operatorname{Vec}(E))$, equivalently the representation category of the Drinfeld double $D(E)$. The condensation has a concrete electric-charge interpretation: the first step condenses the quotient electric charge pulled back from $G$, while the second step condenses the remaining $N$-charge anyons.

\subsubsection{Example 1: the trivial extension $\mathbb{Z}_2\times\mathbb{Z}_2$.}\label{sec:example1}
Let
\begin{equation}
    E=\mathbb{Z}_2\times\mathbb{Z}_2=\{1,a,b,c\},
    \qquad a^2=b^2=1,\quad c=ab=ba.
\end{equation}
We take $\operatorname{Vec}(E)$ as the input of the parent topological order. The corresponding $\mathbb{Z}_2\times\mathbb{Z}_2$ quantum-double phase can be viewed as a stack of two decoupled toric code layers.

For the first step of condensation, $\mathbb{Z}_2\times\mathbb{Z}_2$ can be viewed as the trivial extension
\begin{equation}
    1\longrightarrow N=\{1,a\}
    \longrightarrow E
    \longrightarrow G=E/N\simeq \mathbb{Z}_2
    \longrightarrow 1.
\end{equation}
We identify the quotient classes $N$ and $bN$ with $+$ and $-$, respectively. The quotient $G$-grading of $\operatorname{Vec}(E)$ is
\begin{equation}
    \Vec(\mathbb{Z}_2\times\mathbb{Z}_2)_+=\{1,a\},\qquad
    \Vec(\mathbb{Z}_2\times\mathbb{Z}_2)_-=\{b,c\}.
\end{equation}
The corresponding first multifusion category is
\begin{equation}\label{eq:example1multifusion1}
    \mathcal{M}^{\mathbb{Z}_2}_{\operatorname{Vec}({\mathbb{Z}_2\times\mathbb{Z}_2})}
    =
    \begin{pmatrix}
        \{1_{+,+},a_{+,+}\} & \{b_{+,-},c_{+,-}\}\\
        \{b_{-,+},c_{-,+}\} & \{1_{-,-},a_{-,-}\}
    \end{pmatrix},
\end{equation}
which is the input of a $\Z_2$-symmetry-enriched $\Z_2$ toric code phase. Physically, the first condensation can be viewed as condensing the pure chargeon in one $\Z_2$ toric code layer while leaving the other $\Z_2$ toric code layer untouched. Thus, the child SET phase is a stack of one $\Z_2$ toric code layer and one $\Z_2$-SPT layer, with the two layers remaining decoupled.

For the second-step condensation, the multifusion category in \eqref{eq:example1multifusion1} admits a nontrivial $\mathbb{Z}_2$-grading:
\begin{equation}\label{eq:example1multifusion grade}
    \mathcal{M}^{(+)}
    =
    \{1_{+,+},b_{+,-},b_{-,+},1_{-,-}\},
    \qquad
    \mathcal{M}^{(-)}
    =
    \{a_{+,+},c_{+,-},c_{-,+},a_{-,-}\}.
\end{equation}
The corresponding multifusion category of the grading \eqref{eq:example1multifusion grade} is\footnote{In the multifusion subscripts below, each pair such as $(+,-)$ combines the new $\Z_2$ index from the second grading with the original $\Z_2$ domain index of the intermediate SET. We identify these pairs with elements of $\Z_2\times\Z_2$ when comparing the resulting multifusion category with $\mathcal{M}^{\Z_2\times\Z_2}_{\operatorname{Vec}(\Z_2\times\Z_2)}$.}
\begin{equation}\label{eq:example1multifusion2}
    \mathcal{M}^{\mathbb{Z}_2}_{\mathcal{M}^{\mathbb{Z}_2}_{\operatorname{Vec}({\mathbb{Z}_2\times\mathbb{Z}_2})}}
    =
    \begin{pmatrix}
        1_{(+,+),(+,+)} & b_{(+,+),(+,-)} & a_{(+,+),(-,+)} & c_{(+,+),(-,-)}\\
        b_{(+,-),(+,+)} & 1_{(+,-),(+,-)} & c_{(+,-),(-,+)} & a_{(+,-),(-,-)}\\
        a_{(-,+),(+,+)} & c_{(-,+),(+,-)} & 1_{(-,+),(-,+)} & b_{(-,+),(-,-)}\\
        c_{(-,-),(+,+)} & a_{(-,-),(+,-)} & b_{(-,-),(-,+)} & 1_{(-,-),(-,-)}
    \end{pmatrix}
    \simeq \M^{{\mathbb{Z}_2\times\mathbb{Z}_2}}_{\operatorname{Vec}({\mathbb{Z}_2\times\mathbb{Z}_2})},
\end{equation}
which is the input of the child phase after the second condensation. Physically, this second step can be viewed as condensing the pure chargeon in the remaining toric code layer. Since the $\Z_2$-SPT layer and the toric code layer are decoupled, the condensate is invariant under the $\Z_2$ global symmetry, and hence the global symmetry. The child phase is a stack of two decoupled $\Z_2$-SPT layers, namely a $\Z_2\times\Z_2$-SPT phase, as shown in \eqref{eq:example1multifusion2}.

\subsubsection{Example 2: the non-Abelian extension $S_3=\mathbb{Z}_3\rtimes\mathbb{Z}_2$.}
Let
\begin{equation}
    S_3=\{1,r,r^2,s,rs,sr\},
    \qquad r^3=s^2=1,\quad srs=r^{-1}.
\end{equation}
We take $\operatorname{Vec}(S_3)$ as the input of the parent topological order. The corresponding $S_3$ quantum-double phase has $8$ anyons, conventionally labeled by the conjugacy class of the flux and an irreducible representation of the corresponding centralizer. These anyons are listed in Table \ref{tab:S3 anyon}.
\begin{table}[h]
    \centering
    \begin{tabular}{|c|c|c|c|c|}
        \hline
        \text{Anyon type} & \text{Flux class} & \text{Centralizer} & \text{Irrep of centralizer} & \text{Quantum dimension}\\ 
        \hline
        $A$ & $[1]$ & $S_3$ & $0$ & $1$\\
        $B$ & $[1]$ & $S_3$ & $4$ & $1$\\
        $C$ & $[1]$ & $S_3$ & $2$ & $2$\\
        \hline
        $D$ & $[r]=\{r,r^2\}$ & $\Z_3$ & $1$ & $2$\\
        $E$ & $[r]=\{r,r^2\}$ & $\Z_3$ & $\omega$ & $2$\\
        $F$ & $[r]=\{r,r^2\}$ & $\Z_3$ & $\omega^*$ & $2$\\
        \hline
        $G$ & $[s]=\{s,rs,sr\}$ & $\Z_2$ & $+$ & $3$\\
        $H$ & $[s]=\{s,rs,sr\}$ & $\Z_2$ & $-$ & $3$\\
        \hline
    \end{tabular}
    \caption{Anyons in the $S_3$ quantum-double phase. Here, $0,4$ and $2$ label the trivial representation, sign representation and standard representation of $S_3$, respectively, and $\omega=\exp(2\pi\mathrm{i}/3)$.}
    \label{tab:S3 anyon}
\end{table}

Let's consider the first step of condensation. $S_3$ realizes the short exact sequence
\begin{equation}
    1\longrightarrow N=\{1,r,r^2\}\simeq \mathbb{Z}_3
    \longrightarrow S_3
    \longrightarrow G\simeq \mathbb{Z}_2
    \longrightarrow 1,
\end{equation}
where the nontrivial element of $G=\mathbb{Z}_2$ acts on $N=\mathbb{Z}_3$ by inversion. This gives $\Vec(S_3)$ a $\mathbb{Z}_2$-grading:
\begin{equation}
    \Vec(S_3)_+=\{1,r,r^2\},
    \qquad
    \Vec(S_3)_-=\{s,rs,sr\}.
\end{equation}
The corresponding first multifusion category is
\begin{equation}\label{eq:example2multifusion1}
    \mathcal{M}^{\mathbb{Z}_2}_{\operatorname{Vec}({S_3})}
    =
    \begin{pmatrix}
        \{1_{+,+},r_{+,+},r^2_{+,+}\} & \{s_{+,-},rs_{+,-},sr_{+,-}\}\\
        \{s_{-,+},rs_{-,+},sr_{-,+}\} & \{1_{-,-},r_{-,-},r^2_{-,-}\}
    \end{pmatrix},
\end{equation}
which is the input of a charge-conjugation-symmetry-enriched $\Z_3$ toric code phase. Physically, the first condensation condenses the quotient $\mathbb{Z}_2$ electric charge, namely the sign charge $B$ in Table \ref{tab:S3 anyon}. Since the corresponding sign representation acts trivially only on the normal subgroup $\mathbb{Z}_3=\{1,r,r^2\}$, the $B$ condensation confines any anyon with flux in $\{s,rs,sr\}$, namely $G$ and $H$. The unconfined sector is the quantum double $D(\mathbb{Z}_3)$ of the normal subgroup $\Z_3$. More explicitly, we label the anyons of $D(\mathbb{Z}_3)$ by $(k,\ell)$, where $k\in\mathbb{Z}_3$ is the flux and $\ell\in\mathbb{Z}_3$ is the electric charge. The surviving $D(S_3)$ anyons split as follows:
\[
\begin{aligned}
    A\sim B &\longmapsto (0,0),\\
    C &\longmapsto (0,1)\oplus(0,2),\\
    D &\longmapsto (1,0)\oplus(2,0),\\
    E &\longmapsto (1,1)\oplus(2,2),\\
    F &\longmapsto (1,2)\oplus(2,1),
\end{aligned}
\]
while $G$ and $H$ are confined. Thus, the first condensation produces a $\mathbb{Z}_2$-SET whose local topological order is $D(\mathbb{Z}_3)$. The $\mathbb{Z}_2$ global symmetry acts by quotient conjugation, $r\mapsto r^{-1}$, and hence
\begin{equation}\label{eq:charge conjugation symm}
    (k,\ell)\longmapsto (-k,-\ell).
\end{equation}
For example, it exchanges $(0,1)$ with $(0,2)$, $(1,0)$ with $(2,0)$, $(1,1)$ with $(2,2)$, and $(1,2)$ with $(2,1)$. This global symmetry is known as the charge conjugation symmetry of the $\Z_3$ toric code phase.

For the second-step condensation, the multifusion category in \eqref{eq:example2multifusion1} admits a nontrivial $\mathbb{Z}_3$-grading:
\begin{equation}\label{eq:example2multifusion grade}
    \begin{aligned}
        \mathcal{M}^{(0)}&=\{1_{+,+},s_{+,-},s_{-,+},1_{-,-}\},\\
        \mathcal{M}^{(1)}&=\{r_{+,+},rs_{+,-},sr_{-,+},r^2_{-,-}\},\\
        \mathcal{M}^{(2)}&=\{r^2_{+,+},sr_{+,-},rs_{-,+},r_{-,-}\}.
    \end{aligned}
\end{equation}
The corresponding multifusion category of the grading \eqref{eq:example2multifusion grade} is
\begin{equation}\label{eq:example2multifusion2}
    \begin{aligned}
        \mathcal{M}_{\mathcal{M}_{\text{Vec}( S_{3})}^{\mathbb{Z}_{2}}}^{\mathbb{Z}_{3}} &=\begin{pmatrix}
        1_{( 0,+) ,( 0,+)} & r_{( 0,+) ,( 1,+)} & r_{( 0,+) ,( 2,+)}^{2} & s_{( 0,+) ,( 0,-)} & rs_{( 0,+) ,( 1,-)} & sr_{( 0,+) ,( 2,-)}\\
        r_{( 1,+) ,( 0,+)}^{2} & 1_{( 1,+) ,( 1,+)} & r_{( 1,+) ,( 2,+)} & sr_{( 1,+) ,( 0,-)} & s_{( 1,+) ,( 1,-)} & rs_{( 1,+) ,( 2,-)}\\
        r_{( 2,+) ,( 0,+)} & r_{( 2,+) ,( 1,+)}^{2} & 1_{( 2,+) ,( 2,+)} & rs_{( 2,+) ,( 0,-)} & sr_{( 2,+) ,( 1,-)} & s_{( 2,+) ,( 2,-)}\\
        s_{( 0,-) ,( 0,+)} & sr_{( 0,-) ,( 1,+)} & rs_{( 0,-) ,( 2,+)} & 1_{( 0,-) ,( 0,-)} & r_{( 0,-) ,( 1,-)}^{2} & r_{( 0,-) ,( 2,-)}\\
        rs_{( 1,-) ,( 0,+)} & s_{( 1,-) ,( 1,+)} & sr_{( 1,-) ,( 2,+)} & r_{( 1,-) ,( 0,-)} & 1_{( 1,-) ,( 1,-)} & r_{( 1,-) ,( 2,-)}^{2}\\
        sr_{( 2,-) ,( 0,+)} & rs_{( 2,-) ,( 1,+)} & s_{( 2,-) ,( 2,+)} & r_{( 2,-) ,( 0,-)}^{2} & r_{( 2,-) ,( 1,-)} & 1_{( 2,-) ,( 2,-)}
        \end{pmatrix}\\
        &\simeq\begin{pmatrix}
        1_{1 ,1} & r_{1 ,r} & r_{1 ,r^2}^{2} & s_{1 ,s} & rs_{1 ,rs} & sr_{1 ,sr}\\
        r_{r ,1}^{2} & 1_{r ,r} & r_{r ,r^2} & sr_{r ,s} & s_{r ,rs} & rs_{r ,sr}\\
        r_{r^2 ,1} & r_{r^2 ,r}^{2} & 1_{r^2 ,r^2} & rs_{r^2 ,s} & sr_{r^2 ,rs} & s_{r^2 ,sr}\\
        s_{s ,1} & sr_{s ,r} & rs_{s ,r^2} & 1_{s ,s} & r_{s ,rs}^{2} & r_{s ,sr}\\
        rs_{rs ,1} & s_{rs ,r} & sr_{rs ,r^2} & r_{rs ,s} & 1_{rs ,rs} & r_{rs ,sr}^{2}\\
        sr_{sr ,1} & rs_{sr ,r} & s_{sr ,r^2} & r_{sr ,s}^{2} & r_{sr ,rs} & 1_{sr ,sr}
        \end{pmatrix}=\mathcal{M}_{\text{Vec}( S_{3})}^{S_{3}},
    \end{aligned}
\end{equation}
which is the input of the child phase after the second condensation. Physically, this second condensation condenses the $\mathbb{Z}_3$ electric charge algebra in the $D(\mathbb{Z}_3)$ sector:
\begin{equation}\label{eq:Z3condensate}
    (0,0)\oplus(0,1)\oplus(0,2).
\end{equation}
Since the charge conjugation symmetry in \eqref{eq:charge conjugation symm} permutes $(0,1)$ and $(0,2)$ while leaving $(0,0)$ unchanged, the whole condensate \eqref{eq:Z3condensate} is invariant under the global charge conjugation symmetry. Hence, the second condensation preserves the global symmetry. The child phase is an $S_3$-SPT phase, as shown in \eqref{eq:example2multifusion2}.

\subsubsection{Example 3: the nontrivial central extension $\mathbb{Z}_4$.}
We take $\operatorname{Vec}(\Z_4)=\{0,1,2,3\}$ as the input of the parent topological order. The corresponding $\Z_4$ toric code phase has $16$ anyons labeled by a pair $(k,\ell)$, where $k\in\mathbb{Z}_4$ is the flux and $\ell\in\mathbb{Z}_4$ is the electric charge.

Let's consider the first step of condensation. The group $\Z_4$ can be viewed as a central extension with a nontrivial $2$-cocycle:
\begin{equation}
    1\longrightarrow N=\{0,2\}\simeq \mathbb{Z}_2
    \longrightarrow \mathbb{Z}_4
    \longrightarrow G\simeq \mathbb{Z}_2
    \longrightarrow 1.
\end{equation}
This gives $\Vec(\Z_4)$ a quotient $\mathbb{Z}_2$-grading:
\begin{equation}
    \Vec(\Z_4)_+=\{0,2\},
    \qquad
    \Vec(\Z_4)_-=\{1,3\}.
\end{equation}
The corresponding first multifusion category is
\begin{equation}\label{eq:example3multifusion1}
    \mathcal{M}^{\mathbb{Z}_2}_{\operatorname{Vec}({\mathbb{Z}_4})}
    =
    \begin{pmatrix}
        \{0_{+,+},2_{+,+}\} & \{1_{+,-},3_{+,-}\}\\
        \{1_{-,+},3_{-,+}\} & \{0_{-,-},2_{-,-}\}
    \end{pmatrix},
\end{equation}
which is the input of a $\Z_2$-symmetry-enriched $\Z_2$ toric code phase, obtained by condensing the pure chargeon $(0,2)$ in the $\Z_4$ toric code phase. After condensation, anyons with flux in $\{1,3\}$ are confined, while the surviving anyons are identified as $\Z_2$ toric code anyons:
\[
\begin{aligned}
    (0,0)\sim (0,2) &\longmapsto 1,\\
    (0,1)\sim (0,3) &\longmapsto e,\\
    (2,0)\sim (2,2) &\longmapsto m,\\
    (2,1)\sim (2,3) &\longmapsto f.
\end{aligned}
\]
The global $\Z_2$ symmetry does not permute the anyon species of the resulting $\Z_2$ toric code, similar to the intermediate $\Z_2$-SET \eqref{eq:example1multifusion1} obtained in \textit{Example 1} (Section \ref{sec:example1}). Nevertheless, the global symmetry action is nontrivial in the present case: the anyons $e$ and $f$ acquire a $-1$ phase under the global symmetry transformation, which is known as a symmetry fractionalized charge. This effect is determined by the nontrivial $2$-cocycle in the group extension. Thus, this intermediate $\Z_2$-symmetry-enriched $\Z_2$ toric code phase is distinct from the intermediate SET \eqref{eq:example1multifusion1} obtained in \textit{Example 1}.

Despite the nontrivial symmetry fractionalized charge, this SET still admits symmetry-preserving anyon condensation. To see this, note that the input multifusion category \eqref{eq:example3multifusion1} admits a $\Z_2$-grading:
\begin{equation}
    \mathcal{M}^{(+)}
    =
    \{0_{+,+},1_{+,-},3_{-,+},0_{-,-}\},\qquad
    \mathcal{M}^{(-)}
    =
    \{2_{+,+},3_{+,-},1_{-,+},2_{-,-}\},
\end{equation}
from which we can construct the second multifusion category
\begin{equation}
    \mathcal{M}^{\mathbb{Z}_2}_{\mathcal{M}^{\mathbb{Z}_2}_{\operatorname{Vec}({\mathbb{Z}_4})}}
    =
    \begin{pmatrix}
        0_{(+,+),(+,+)} & 1_{(+,+),(+,-)} & 2_{(+,+),(-,+)} & 3_{(+,+),(-,-)}\\
        3_{(+,-),(+,+)} & 0_{(+,-),(+,-)} & 1_{(+,-),(-,+)} & 2_{(+,-),(-,-)}\\
        2_{(-,+),(+,+)} & 3_{(-,+),(+,-)} & 0_{(-,+),(-,+)} & 1_{(-,+),(-,-)}\\
        1_{(-,-),(+,+)} & 2_{(-,-),(+,-)} & 3_{(-,-),(-,+)} & 0_{(-,-),(-,-)}
    \end{pmatrix}\simeq\mathcal{M}^{\mathbb{Z}_4}_{\operatorname{Vec}({\mathbb{Z}_4})},
\end{equation}
as the input of the child phase after the second condensation. The existence of the grading above, together with this multifusion-category construction, shows that the second condensation is compatible with the global symmetry and therefore preserves it.

Physically, the second condensation corresponds to condensing the pure chargeon $e$ in the intermediate $\Z_2$-SET \eqref{eq:example3multifusion1}. One might expect this condensation to break the global symmetry, since $e$ acquires a $-1$ phase under the symmetry and hence the condensate $1+e$ does not appear invariant; this is precisely the conclusion reached from the abstract condensable-algebra point of view in Ref.\cite{bischoff2019spontaneous}. Our input-level analysis instead concerns physical Hilbert-space states. In the coherent-state picture of anyon condensation, the condensed vacuum is built as a coherent state associated with creation operators for the condensate anyons. Since $e$ anyons must be created in pairs in a physical state, the relevant condensate state remains invariant under the global symmetry. Thus, the second condensation preserves global symmetry, in agreement with the multifusion construction.

\section{Discussion}

In this work, we formulated symmetry-preserving anyon condensation in SET phases directly at the level of the multifusion-category input of the enlarged HGW string-net model. The central criterion is the existence of a compatible grading of the input multifusion category.

More concretely, for a $G$-SET phase with input $\M^G_{\Fus}$, a nontrivial $N$-grading of $\M^G_{\Fus}$ provides the input-level criterion for a $G$-preserving condensation. The child phase is again described by a multifusion category, namely $\M^{N}_{\M^G_{\Fus}}$. For an $E$-graded parent fusion category with $1\to N\to E\to G\to 1$, the two-step construction gives
\begin{equation}
    \mathcal{M}^{N}_{\mathcal{M}^{G}_{\mathscr{F}}}
    \simeq
    \mathcal{M}^{E}_{\mathscr{F}},
\end{equation}
while keeping the intermediate $G$-SET phase and the symmetry-preserving nature of the second condensation explicit.

The examples illustrate how this criterion works in practice. The EM-exchange-enriched toric code shows that a grading of the remaining diagonal component is not sufficient; the grading must also be compatible with the domain-wall sectors, the off-diagonal components. The three quantum-double examples exhibit allowed cases for a trivial extension, a non-Abelian semidirect product, and a nontrivial central extension. In particular, the $\mathbb{Z}_4$ example shows that nontrivial symmetry fractionalization does not by itself forbid condensation once the condensate is treated at the level of physical states.

Several directions remain open. The most immediate one is to extend the present framework beyond pure chargeon condensation to general dyon condensation \cite{zhao2025nonabelian}. Such condensations are not expected to be described simply by grading the input category because dyonic condensates involve flux-charge data and do not arise merely by gapping degrees of freedom associated with a group grading. A second natural direction is to generalize the present group-symmetry framework to category symmetries, or more broadly generalized symmetries.

\begin{acknowledgments}
The authors thank Yuting Hu, Yuxiang Wang, Yifei Wang for inspiring and helpful discussions. YW is supported by NSFC Grant No. 12475001, the Shanghai Municipal Science and Technology Major Project (Grant No. 2019SHZDZX01), Science and Technology Commission of Shanghai Municipality (Grant No. 24LZ1400100), and the Quantum Science and Technology-National Science and Technology Major Project (No. 2024ZD0300101). YW is an affiliate member of the Institute for Quantum Computing. YW is grateful for the hospitality of the Perimeter Institute during his visit, where the main part of this work is done. This research was supported in part by the Perimeter Institute for Theoretical Physics. Research at Perimeter Institute is supported by the Government of Canada through the Department of Innovation, Science and Economic Development and by the Province of Ontario through the Ministry of Research, Innovation and Science. 
\end{acknowledgments}

\appendix

\section{String-net Model}\label{app:review}

\begin{figure}[!ht]
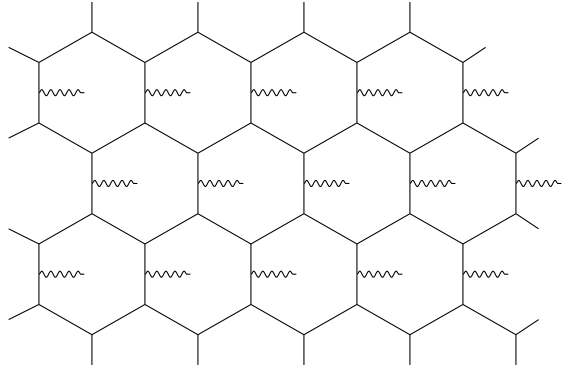

\centering
\Lattice
\caption{HGW string-net model lattice. Each plaquette has a tail (wavy line).}
\label{fig:lat}
\end{figure}

This section reviews the string-net model of Ref.\cite{zhao2022}, adapted from the construction in Ref.\cite{Hu2018}. The model is exactly solvable on a \(2\)-dimensional lattice; one example is shown in Fig. \ref{fig:lat}. We take all vertices to be trivalent. Each plaquette carries an inward-pointing tail attached to one of its edges. Changing the attachment edge of a tail gives an equivalent description, as to be explained in Appendix \ref{sec:pachner}. Edges and tails are oriented, and reversing their orientations is likewise an equivalent choice.

The input of the model is a unitary fusion category \(\Fus\). We describe it by a finite set \(L_\Fus\) of \emph{simple objects}, together with three functions \(N: L_\Fus^3 \to \NN\), \(d: L_\Fus \to \RR^+\), and \(G: L^6_\Fus \to \C\). The function \(N\) specifies the \emph{fusion rules} and obeys
\eq{
\sum_{e \in L_\Fus} N_{ab}^e N_{ec}^d = \sum_{f \in L_\Fus} N_{af}^d N_{bc}^f, \qquad\qquad N_{ab}^c = N_{c^\ast a}^{b^\ast}.
}
There is a distinguished simple object \(1 \in L_\Fus\), the \emph{trivial object}, such that for any \(a, b \in L_\Fus\),
\eq{
N_{1a}^b = N_{1b}^a = \delta_{ab},
}
where \(\delta\) denotes the Kronecker symbol. Each \(a \in L_\Fus\) has a unique \emph{opposite object} \(a^\ast \in L_\Fus\), characterized by
\eq{
N_{ab}^1 = N_{ba}^1 = \delta_{ba^\ast}.
}
Throughout this appendix, we restrict to the multiplicity-free case, where \(N_{ab}^c\) is either \(0\) or \(1\) for all \(a, b, c \in L_\Fus\). We then define
\eq{
\delta_{abc} = N_{ab}^{c^\ast} \in \{0, 1\}.
}

The basis configurations are obtained by assigning a simple object in \(L_\Fus\) to every edge and tail. At each vertex, the three incident edges or tails are taken to point toward the vertex and are labeled counterclockwise by \(i, j, k \in L_\Fus\); the allowed configurations satisfy \(\delta_{ijk}=1\). Reversing the orientation of an edge or tail while replacing its label by the opposite object, \(j\to j^\ast\), leaves the configuration unchanged. The Hilbert space \(\Hil\) is spanned by all such admissible labelings.

The function \(d: L_\Fus\to\RR^+\) gives the \emph{quantum dimensions} of the simple objects. These dimensions are the largest eigenvalues of the fusion matrices and form a \(1\)-dimensional representation of the fusion rules:
\eq{d_ad_b = \sum_{c\in L_\Fus}N_{ab}^cd_c.}
In particular, \(d_1 = 1\), and \(d_a = d_{a^\ast}\ge 1\) for every \(a\in L_\Fus\). 

The function \(G: L_\Fus^6\to\CC\) gives the \(6j\)-\emph{symbols} of the fusion algebra and satisfies
\eqn[eq:sixj]{\sum_nd_nG^{pqn}_{v^*u^*a}G^{uvn}_{j^*i^*b}G^{ijn}_{q^*p^*c} = G^{abc}_{i^*pu^*}&G^{c^*b^*a^*}_{vq^*j},\qquad\sum_nd_nG^{ijp}_{kln}G^{j^*i^*q}_{l^*k^*n} = \frac{\delta_{pq^*}}{d_p}\delta_{ijp}\delta_{klq},\\
G^{ijm}_{kln} = G^{klm^*}_{ijn^*} = G^{jim}_{lkn^*}= G^{mij}_{nk^*l^*} &= \alpha_m\alpha_n\overline{G^{j^*i^*m^*}_{l^*k^*n^*}},\qquad \Big|G^{abc}_{1bc}\Big| = \frac{1}{\sqrt{d_bd_c}}\delta_{abc},
}
where \(\alpha_m = G^{1mm^\ast}_{1m^\ast m}\in\{\pm 1\}\) is the Frobenius-Schur indicator of the simple object \(m\).

The string-net Hamiltonian is
\eqn[eq:hamiltonian]{H := - \sum_{{\rm Plaquettes}\ P}Q_P,\qquad\qquad Q_P := \frac{1}{D}\sum_{s\in L_\Fus}d_sQ_P^s,}
where \(Q_P^s\) acts on the edges around plaquette \(P\). On a hexagonal plaquette, its matrix elements are
\begin{align*}Q_P^s\ \PlaquetteSrc\
:=\ \delta_{p,1}\delta_{j_1,j_7}\ \sum_{j_k\in L_\Fus}\ \prod_{k = 1}^{6}\ \Bigg(\sqrt{d_{i_k}d_{j_k}}G^{e_ki_ki_{k+1}^\ast}_{sj_{k+1}^\ast j_k}\Bigg)\PlaquetteTar\ ,
\end{align*}
and
\eq{D := \sum_{a\in L_\Fus}d_a^2}
is the total quantum dimension of the UFC \(\Fus\). We display only the action of \(Q_P\) on a hexagonal plaquette; the matrix elements for other plaquette shapes are defined analogously. For compactness, the ``$\ket{\cdot}$'' labels around diagrams are omitted unless needed.

These operators satisfy
\eq{
(Q_P^s)^\dagger = Q_P^{s^\ast},\qquad Q_P^rQ_P^s = \sum_{t\in L_\Fus} N_{rs}^tQ_P^t,\qquad Q_P^2 = Q_P,\qquad Q_{P_1}Q_{P_2} = Q_{P_2}Q_{P_1}.
}
The terms \(Q_P\) in \(H\) are commuting projectors, making the Hamiltonian exactly solvable. The ground-state subspace \(\Hil_0\) is obtained by the projection
\eqn{
\Hil_0 = \Bigg[\prod_{{\rm Plaquettes\ } P}Q_P\Bigg]\Hil.
}
On a spherical lattice, the ground state \(\ket\Phi\) is unique up to an overall scalar.

\subsection{Topological Features}\label{sec:pachner}

We next recall the topological invariance of the ground-state subspace in the string-net model of Ref.\cite{Hu2018}. Any two lattices with the same topology are related by \emph{Pachner moves}. When the input UFC is fixed, lattices connected by such moves have Hilbert spaces related by unitary linear maps \cite{Hu2012}, denoted by \(\T\), and the ground states are invariant under these maps. The three elementary Pachner moves are represented by the following transformations:
\eqn[eq:pachner]{
&\T \quad \PachnerOne\ ,\\
&\T \quad \PachnerTwo\ ,\\
&\T \quad \PachnerThree\ .}
Here, ``\({\color{red}\times}\)'' marks the plaquette being contracted. These elementary moves, together with their unitary transformations, can be composed. Although a pair of initial and final lattices can be connected by different sequences of Pachner moves, all such sequences induce the same transformation matrix on the ground-state Hilbert space.

As noted above, the edge of a plaquette to which the tail is attached is not part of the topological data. Different choices produce different lattice configurations, and hence different Hilbert-space presentations, but they are related by the linear transformation \(\T'\):
\eqn[eq:tailmove]{
\T'\quad\PachnerFour\ .
}
Configurations with tails attached to other edges are obtained by iterating this move.

In some intermediate manipulations, it is convenient to allow auxiliary states with more than one tail in a plaquette. Such states are not new physical states: they are equivalent to ordinary Hilbert-space states with one tail per plaquette through
\eqn[eq:tailfuse]{
\PachnerFive\ .
}

\subsection{Excited States}\label{sec:spec}

An \emph{excited state} \(\ket\varphi\) is an eigenstate for which \(Q_P\ket\varphi = 0\) on at least one plaquette \(P\). Such plaquettes are said to host \emph{anyons}. Ground states may be regarded as trivial excited states, with only \emph{trivial anyons} on every plaquette. We work with sphere topology, where the ground state is unique, although the construction below applies to other topologies as well.

We begin with the simplest excitation: a pair of anyons placed in two \emph{adjacent} plaquettes sharing an edge \(E\). It is created by a ribbon operator \(W_E^{J;pq}\), obtained by composing an auxiliary operator \(\mathcal{W}_E^{J; pq}\)
\eqn[eq:create]{
\mathcal{W}_E^{J; pq} \ExcitedA := \sum_{k \in L_\Fus} \sqrt{\frac{d_k}{d_{j_E}}} \ {z_{pqj_E}^{J;k}} \ \ \ExcitedB \ 
}
with the Pachner moves \eqref{eq:tailmove} and \eqref{eq:tailfuse}, where $j_E$ denotes the dof on edge $E$. In the original Hilbert space, every plaquette has exactly one tail attached to a chosen edge, which may be \(E\). The auxiliary operator \(\mathcal{W}_E^{J; pq}\) in \eqref{eq:create} creates two additional tails on \(E\); the remaining Pachner moves in the creation operator then move and fuse these tails with the original plaquette tails.

The coefficients \(z_{pqj}^{J; k}\) appearing in Eq.\eqref{eq:create} form the \emph{half-braiding tensor} of anyon type \(J\). They are determined by
\eqn[eq:halfA]{
\frac{\delta_{jt}N_{rs}^t}{d_t} z_{pqt}^{J;w} = \sum_{u,l,v\in L_\Fus} z_{lqr}^{J;v} z_{pls}^{J;u}d_u d_v G^{r^*s^*t}_{p^*wu^*} G^{srj^*}_{qw^*v} G^{s^*ul^*}_{rv^*w}.
}
Appendix \ref{sec:halfbraid} explains the physical meaning of this equation. The label \(J\), the \emph{anyon type}, indexes the minimal solutions \(z^J\) of Eq.\eqref{eq:halfA}, meaning solutions that cannot be decomposed as sums of nonzero solutions. Categorically, anyon types are simple objects in the \emph{center} \(\Cent(\Fus)\) of the UFC \(\Fus\). This center is a modular tensor category whose data encode the topological order realized by the string-net model. In particular, the trivial anyon $\mathcal{I}\in L_{\Cent(\Fus)}$ satisfies
$$z^{\mathcal{I};k}_{pqj} = \delta_{p1}\delta_{q1}\delta_{jk}.$$

The statistics of an anyon \(J\) are given by the corresponding diagonal entry of the modular \(T\) matrix of the UMTC \(\Cent(\Fus)\):
\eq{
T_{JK} = \frac{1}{d_t}\delta_{JK}\sum_{p \in L_\Fus} d_p z_{ttt}^{J;p}.}
Here, \(t\) is any flux carried by \(J\). The mutual braiding of anyons \(J\) and \(K\) is encoded in the modular \(S\) matrix, with entries
\eq{S_{JK} = \frac{1}{D}\ \sum_{p, q, k \in L_\Fus} d_k \bar{z}_{ppq}^{J;k} \bar{z}_{qqp}^{K;k}.
}
When \(\theta_J = 1\), we say anyon \(J\) has trivial self-statistics. Two anyons \(J\) and \(K\) braid trivially if and only if \(S_{JK} = d_J d_K\), where \(d_J\) is the quantum dimension of \(J\), defined by
\eq{
d_J = \sum_{J\text{'Fluxes } p} d_p.
}

For non-adjacent plaquettes, a pair of quasiparticles is created by a ribbon operator supported on a longer path, obtained by concatenating the elementary ribbon operators. For instance, to create \(J^\ast\) and \(J\) with fluxes \(p^\ast_0\) and \(p_n\) in plaquettes \(P_0\) and \(P_n\), choose a plaquette sequence \((P_0, P_1, \cdots, P_n)\) such that \(P_i\) and \(P_{i+1}\) share an edge \(E_i\). The corresponding ribbon operator is
\eq{
W_{P_0P_n}^{J; p_0p_n} := \Bigg[\sum_{p_1 p_2 \cdots p_{n-1} \in L_\Fus} \prod_{k = 1}^{n-1} \left(d_{p_k} B_{P_k} W_{E_k}^{J; p_k p_{k+1}}\right)\Bigg]W_{E_0}^{J; p_0 p_1}.
}
If two such plaquette paths can be continuously deformed into one another, they define the same operator \(W_{P_0P_n}^{J; p_0 p_n}\). The same construction extends to creation operators for three or more anyons.

Finally, the measurement operator \(M_P^J\), which detects whether plaquette \(P\) carries an anyon \(J\), is
\eqn{M_P^J\ \PlaquetteSrc
:= \sum_{s,t\in L_\Fus} \frac{d_sd_tz_{pps}^{J; t}}{d_p} \PlaquetteMsr\ .}
These measurement operators are orthonormal and complete:
\eq{M_P^JM_P^K = \delta_{JK}M_P^J,\qquad \sum_{J\in L_{\mathcal{Z}(\Fus)}}M_P^J = \idm.}

\subsection{The Output UMTC is the Center of the Input UFC}\label{sec:halfbraid}

As reviewed above, the output UMTC of the string-net model is the center \(\Cent(\Fus)\) of the input UFC \(\Fus\). The lattice construction provides a concrete realization of this center relation, which we now spell out.

Categorically, an object \(J\) in \(\Cent(\Fus)\) is a pair \(J = (X_J, c_{X_J, \cdot})\). Here \(X_J\) is an object of the UFC \(\Fus\), and \(c_{X_J, \cdot}\), the \emph{half-braiding}, is a collection of morphisms
\eq{
\{ c_{X_J, y}: X_J\otimes y\to y\otimes X_J | y\in \Fus
\}. }
The morphism \(c_{X_J, y}\) braids \(X_J\) with \(y\) inside \(\Fus\), diagrammatically
\eq{\HalfBraidingA.}
Since morphisms in a UFC decompose into channels built from simple-object fusion, the half-braiding can be expanded as
\eqn[eq:halfB]{\HalfBraidingB.}
Here \(L_J\subseteq L_\Fus\) is the subset whose simple objects sum to \(X_J\):
\eq{X_J = \bigoplus_{p \in L_J} p.}
The coefficients \(z_{pqk}^{J;y}\) are the \emph{half-braiding tensor} of \(J\). Compatibility with fusion in \(\Fus\) requires
\eqn[eq:halfC]{\HalfBraidingC.}
Expanding Eq. \eqref{eq:halfC} by means of Eq. \eqref{eq:halfB} gives Eq. \eqref{eq:halfA}.

For a string-net model with input UFC $\Fus$, an anyon type \(J\) is therefore a simple object of the output UMTC \(\Cent(\Fus)\), and \(J\)'s fluxes lie in \(L_J\). The creation operator \(W_E^{J;pq}\) realizes the half-braiding morphism \(c_{X_J, j_E}\) between \(X_J\) and the edge dof \(j_E\in L_\Fus\):
\eq{\HalfBraidingD}

\subsection{The String-Net Model with Non-Commutative Input UFCs}\label{appendix:noab}

We now briefly discuss the string-net model whose input UFC $\Fus$ has non-commutative fusion rules $\delta$, meaning that for some $a, b, c\in L_\Fus$ one has $\delta_{abc}\ne\delta_{acb}$. In this case, a Hilbert-space state must satisfy the vertex constraint \(\delta_{abc} \ne 0\) whenever three incident edges or tails meet at a vertex, are all oriented toward that vertex, and carry degrees of freedom \(a, b, c\) in \emph{counterclockwise} order. 

For non-commutative string-net models, the half-braiding equation \eqref{eq:halfA} may have matrix-valued solutions. Thus, a minimal solution \( z_{pqj}^{J;k} \) need not be a complex number; it can instead be a unitary matrix. Here \(J\) labels the anyon types associated with distinct minimal solutions, and \(p, q, j, k\in L_\Fus\). The tensor \(z^J\) is then written as
\eq{[z_{pqj}^{J;k}]_{\alpha\beta} \in \mathbb{C}, \qquad 1 \leq \alpha \leq n_{(J, p)},\qquad 1\le \beta \leq n_{(J, q)},}
where $n_{(J, p)}\in\mathbb{N}$. When an anyon \(J\) has \(n_{(J,p)}>1\) for a flux \(p\), a dyon type is specified not only by the anyon species \(J\), but also by a multiplicity index \(1 \le \alpha\le n_{(J,p)}\):
$$(J, p, \alpha),\qquad J\in L_{\Cent(\Fus)},\qquad p\text{ is }J\text{'s flux},\qquad 1\le\alpha\le n_{(J, p)}.$$
We call the pair \((J, p)\), consisting of the anyon type \(J\) and flux type \(p\), a \emph{dyon multiplet}. It contains \(n_{(J,p)}\) distinct dyon types \((J,p,\alpha)\), with \(\alpha\) called the \emph{multiplet index}. The integer \(n_{(J,p)}\) is the \emph{degeneracy} of the dyon multiplet \((J,p)\). The pair \((p,\alpha)\), made of the flux type and the multiplet index, is referred to as the \emph{local dof} of the anyon \(J\).

In the Hilbert space of a non-commutative string-net model, two excited states with a dyon in the same plaquette, the same anyon type \(J\), and the same flux type \(p\), but different multiplet indices \(\alpha\), should be orthogonal. The original tail Hilbert space, spanned only by flux labels \(p\in L_\Fus\), is therefore too small to resolve the full dyon spectrum. Following Ref.\cite{zhao2025noninvertible,zhao2025c}, we \emph{enlarge} the local Hilbert space on each \emph{tail}. Edge dofs do not need the same extension: edges belong to ground-state configurations, where paths along edges form closed loops and vertices enforce only the fusion rules, without reference to multiplet indices.

Accordingly, the local subspace on a tail is spanned by dyon local dofs \( p_{J, \alpha} \), where \(J\) is an anyon species carrying flux \(p\), and \(1 \leq \alpha \leq n_{(J,p)}\). Distinct tail basis states are orthogonal:
$$\Bigg\langle\quad\Tail{p_{J,\alpha}}\ \Bigg|\quad\Tail{q_{K,\beta}}\ \Bigg\rangle = \delta_{JK}\delta_{pq}\delta_{\alpha\beta}.$$
We use this enlarged tail Hilbert space even for commutative string-net models. This convention does not change the discussion, because excited states with different anyon types in the same plaquette are orthogonal.

The elementary creation operator $W_E^{J;(p, \alpha)(q, \beta)}$, which creates a pair of dyons $(J^\ast, p^\ast, \alpha)$ and $(J, q, \beta)$ in two adjacent plaquettes separated by edge $E$, is defined by
\eqn{
W_E^{J; (p,\alpha)(q,\beta)} \NonAbA := \sum_{k \in L_\Fus} \sqrt{\frac{d_k}{d_j}}\ \ [{z_{pqj}^{J;k}}]_{\alpha\beta} \ \ \NonAbB \ ,
}
Here, $\mathcal{I}\in L_{\Cent(\Fus)}$ is the trivial anyon, with unique flux $1\in L_\Fus$ and no degeneracy, \(n_\mathcal{I}=1\). Products of creation operators depend on the CG coefficients for tensor products of half-braiding \(z\)-tensors; we do not need their explicit form here.

A standard example of a non-commutative input UFC is \({\rm Vec}(G)\) for a non-Abelian group \(G\). Its simple objects are labeled by group elements, and for any $a, b, c, d, e, f\in G$,
$$\delta_{abc} = 1\quad\text{if}\quad c^\ast = ab\quad\text{else}\quad 0,\qquad\qquad d_a = 1,\qquad\qquad G^{abe}_{cdf} = \delta_{abe}\delta_{bcf^\ast}\delta_{cde^\ast}\delta_{daf},$$
where $c^\ast$ is the inverse of $c\in G$. In this case, Eq. \eqref{eq:halfA} for the \(z\)-tensors becomes
$$z^{J;(pa)}_{p(a^\ast pa)a}z^{J;(a^\ast pab)}_{(a^\ast pa)(b^\ast a^\ast pab)b} = z^{J;(pab)}_{p(b^\ast a^\ast pab)(ab)}.$$
Consequently, an anyon type \(J\) is labeled by a pair
$$J = (\bar p, \rho),$$
where $\bar p = \{g^\ast pg| g\in G\}$ is the conjugacy class of \(G\) containing \(p\), and \(\rho\) is an irreducible representation of the centralizer of \(\bar p\):
$$Z(\bar p) = \{g\in G| g^\ast pg = p\}.$$
The degeneracy of the dyon multiplet \((J,p)\) is the dimension of the irrep \(\rho\) of \(Z(\bar p)\):
$$n_{((\bar p, \rho), p)} = {\rm dim}\rho,$$
and the quantum dimension of \(J\) is \(n_{((\bar p, \rho), p)}|\bar p|\), namely the total number of local dofs associated with \(J\).

In particular, chargeons in the string-net model with input UFC ${\rm Vec}(G)$ are dyons of type \((\bar 1,\rho)\), where the trivial conjugacy class is $\bar 1 = \{1\}\subset G$ and $Z(\bar 1) = G$. The corresponding half-braiding tensor is
$$z_{1,1,r}^{(\bar 1, \rho);r} = D_{\alpha\beta}^\rho(g),\qquad 1\le\alpha, \beta\le{\rm dim}\rho,\qquad\forall r\in G.$$
Here, $D^\rho$ is the representation matrix of the irrep \(\rho\) of the group \(G\).

\section{Multifusion Category as the Input of SET Model}\label{eq:multifusion input}

This section explains how a multifusion category serves as the input data of an SET string-net model.

\subsection{Multifusion Category}\label{app:multifusion}

We first recall the basic structure of a multifusion category. A \textit{multifusion category} $\mathcal{M}$ is a $k$-linear semisimple rigid monoidal category with finitely many simple objects. In contrast to an ordinary fusion category, the tensor unit $\mathbb{1}$ of $\mathcal{M}$ need not be simple. It decomposes into mutually orthogonal simple summands:
\begin{equation}
    \mathbb{1} \simeq \bigoplus_{i\in I}\mathbb{1}_i,
\end{equation}
where \(I\) is a finite index set. The simple sub-units \(\mathbb{1}_i\) behave as orthogonal projectors under tensor product:
\begin{equation}\label{eq:7}
    \mathbb{1}_i \otimes \mathbb{1}_j \simeq \delta_{ij} \mathbb{1}_i.
\end{equation}
Here, \(0\) denotes the zero object \(\mathbb{0}\), which absorbs any object \(X\) under tensor product, \(\mathbb{0} \otimes X \simeq X \otimes \mathbb{0} \simeq \mathbb{0}\), and has trivial morphism spaces.

The unit decomposition induces a block decomposition of the multifusion category into components \(\mathcal{M}_{ij}\):
\begin{equation}\label{eq:multifusion decomposition}
    \mathcal{M} \simeq \bigoplus_{i,j\in I}\mathcal{M}_{ij}, \quad \text{where} \quad \mathcal{M}_{ij} = \mathbb{1}_i \otimes \mathcal{M} \otimes \mathbb{1}_j.
\end{equation}
Accordingly, any object \(X\in\mathcal{M}\) decomposes into block components:
\begin{equation}
    X \simeq \bigoplus_{i,j\in I}(\mathbb{1}_i \otimes X \otimes \mathbb{1}_j) = \bigoplus_{i,j\in I}X_{ij},
\end{equation}
with \(X_{ij}\in\mathcal{M}_{ij}\). The fusion rules \eqref{eq:7} imply that an object in \(\mathcal{M}_{ij}\) absorbs the corresponding left and right units:
\begin{equation}\label{eq:11}
    \mathbb{1}_i \otimes X_{ij} \simeq X_{ij} \otimes \mathbb{1}_j \simeq X_{ij}.
\end{equation}
The tensor product between blocks is nonzero only when the adjacent indices match. For \(X_{ij}\in\mathcal{M}_{ij}\) and \(Y_{kl}\in\mathcal{M}_{kl}\),
\begin{equation}\label{eq:multifusion fusion rules}
    X_{ij} \otimes Y_{kl} \simeq \mathbb{0} \quad \text{if } j \ne k,
\end{equation}
and
\begin{equation}
    X_{ij} \otimes Y_{jk} \simeq \mathbb{1}_i \otimes (X_{ij} \otimes Y_{jk}) \otimes \mathbb{1}_k \in \mathcal{M}_{ik}.
\end{equation}
Rigidity and semisimplicity ensure that \(X_{ij}\otimes Y_{jk}\ne \mathbb{0}\) whenever \(X_{ij}\) and \(Y_{jk}\) are nonzero objects.

The decomposition \eqref{eq:multifusion decomposition} can be displayed as a multifusion matrix:
\begin{equation}\label{eq:multifusionmatrix}
    \M=\begin{pmatrix}
        \M_{11} & \M_{12} & \cdots & \M_{1n} \\
        \M_{21} & \M_{22} & \cdots & \M_{2n} \\
        \vdots & \vdots & \ddots & \vdots \\
        \M_{n1} & \M_{n2} & \cdots & \M_{nn}
    \end{pmatrix}.
\end{equation}
Equation \eqref{eq:multifusion fusion rules} may then be read as the matrix-product rule for this multifusion matrix.

For a diagonal block \(i=j\), the object \(\mathbb{1}_i\in\mathcal{M}_{ii}\) is a simple tensor unit and \(\mathcal{M}_{ii}\times\mathcal{M}_{ii}\to\mathcal{M}_{ii}\). Hence each diagonal component \(\mathcal{M}_{ii}\) is an ordinary fusion category. By contrast, an off-diagonal component \(\mathcal{M}_{ij}\) with \(i\ne j\) is not itself a fusion category; it is an \(\mathcal{M}_{ii}\)-\(\mathcal{M}_{jj}\) bimodule category.

\subsection{A Multifusion Category as the Input Data}

We now explain how multifusion categories enter the lattice description of SET phases.

Pairing a topological phase \(\mathscr{F}\) with a global symmetry \(G\) gives an SET phase, denoted by \(\mathscr{F}_G\). As reviewed in Section \ref{sec:background}, such a phase contains several \(G\)-domains. Each domain is locally equivalent to \(\mathscr{F}\), while different domains are distinguished by elements of the symmetry group. The \(G\) action relates these domains. This domain structure motivates a lattice model for \(\mathscr{F}_G\) whose input is a multifusion category \(\M\) equipped with a \(G\)-action.

The fusion rule \eqref{eq:multifusion fusion rules} suggests thickening the lattice in Fig. \ref{fig:lat}: each edge or tail is replaced by a double line, and group-element indices live on these double lines \cite{Chang2015}, as illustrated in Fig. \ref{fig:fatlattice}. Then each plaquette is assigned a unique group element \(g\), given by the label on the inner loop surrounding that plaquette. We say that the plaquette lies in the \(g\)-domain. Equivalently, this construction can be understood as putting spin variables on the plaquettes \cite{heinrich2016symmetry}.

The simple objects in the diagonal block \(\Fus_{gg}\) of the multifusion matrix \eqref{eq:multifusionmatrix} describe degrees of freedom inside the \(g\)-domain. Off-diagonal blocks instead describe domain-wall degrees of freedom: objects in \(\Fus_{g_ig_j}\) live on domain walls separating a \(g_i\)-domain from a \(g_j\)-domain.

With this input data, the lattice for the SET model of \(\mathscr{F}_G\) has the same underlying form as the HGW string-net lattice for \(\mathscr{F}\): a honeycomb lattice in which every plaquette carries a tail, as shown in Fig. \ref{fig:lat}. Each edge and tail is assigned a degree of freedom valued in the simple objects of \(\M\), subject to the fusion rules of \(\M\) at every vertex.

The Hamiltonian takes the form
\eqn{
H=-\sum_{p}Q_p=-\sum_{p}\sum_{s}Q_p^{s},
}
where \(p\) runs over plaquettes and \(s\) runs over all simple objects of the multifusion category \(\mathcal{M}\). This Hamiltonian has the same formal structure as the HGW string-net Hamiltonian \eqref{eq:hamiltonian} for a pure topological phase, despite now dofs take value in a multifusion category instead of a fusion category.

\section{\textit{G}-graded fusion category}\label{app:g-grading}

This section gives a detailed definition of a $G$-grading of a fusion category. A \textit{$G$-grading} on $\mathscr{F}$ is a direct sum decomposition into \textit{homogeneous components} $\mathscr{F}_g$, each of which is a $k$-linear semisimple category:
\begin{equation}\label{eq:1}
    \mathscr{F} \simeq \bigoplus_{g\in G}\mathscr{F}_g.
\end{equation}
There are no nonzero morphisms between distinct homogeneous components: for \(X\in\mathscr{F}_g\) and \(Y\in\mathscr{F}_h\), one has \(\operatorname{Hom}_\mathscr{F}(X,Y)=0\) whenever \(g\ne h\). Objects lying entirely in a single component \(\mathscr{F}_g\) are called \textit{homogeneous objects}. Hence every object \(X\in\mathscr{F}\) decomposes uniquely as a direct sum of homogeneous objects:
\begin{equation}
    X \simeq \bigoplus_{g\in G}X_g, \quad X_g \in \mathscr{F}_g.
\end{equation}
A \textit{$G$-graded fusion category} is a fusion category $\mathscr{F}$ with a $G$-grading whose monoidal tensor product respects the group multiplication, sending $\mathscr{F}_g \times \mathscr{F}_h$ into $\mathscr{F}_{gh}$. That is,
\begin{equation}
    X_g \otimes Y_h \in \mathscr{F}_{gh}
\end{equation}
for all homogeneous objects $X_g \in \mathscr{F}_g$ and $Y_h \in \mathscr{F}_h$. Although $\mathscr{F}$ itself is a fusion category, the components $\mathscr{F}_g$ with \(g\ne e\) are generally not fusion categories, since they are not closed under tensor product; they are only $k$-linear semisimple categories. 

Every simple object of $\mathscr{F}$ is homogeneous. If \(\operatorname{Irr}(\mathscr{F})\) denotes the set of isomorphism classes of simple objects, the grading defines a degree map \(\operatorname{deg}_G: \operatorname{Irr}(\mathscr{F}) \to G\), with \(\operatorname{deg}_G(a)=g\) precisely when \(a \in \operatorname{Irr}(\mathscr{F}_g)\). Thus, $\cup_{g\in G}\operatorname{Irr}(\mathscr{F}_g) = \operatorname{Irr}(\mathscr{F})$. 

Given a $G$-grading on $\mathscr{F}$, any group homomorphism $\phi: G \to H$ induces an $H$-grading on $\mathscr{F}$:
\begin{equation}\label{eq:5}
    \mathscr{F} \simeq \bigoplus_{h\in H}\mathscr{F}_h, \quad \text{where} \quad \mathscr{F}_h = \bigoplus_{g\in\phi^{-1}(h)}\mathscr{F}_g.
\end{equation}
Equivalently, $\operatorname{deg}_H = \phi \circ \operatorname{deg}_G$. Thus, group homomorphisms control how one passes between gradings. Because \(\phi\) preserves multiplication, the decomposition in \eqref{eq:5} satisfies the condition for an \(H\)-graded fusion category.

Given a fusion category $\mathscr{F}$ with a nontrivial $G$-grading, we construct a corresponding \textit{multifusion category} $\mathcal{M}^G_\mathscr{F}$ by using group elements as block indices:
\begin{equation}
    \mathcal{M}_\mathscr{F}^G \simeq \bigoplus_{g,h\in G}\mathcal{M}_{g,h}, \quad \text{where} \quad \mathcal{M}_{g,h} := \mathscr{F}_{g^{-1}h}.
\end{equation}
The tensor product of homogeneous objects follows the multiplication law encoded by the $G$-grading:
\begin{equation}
    X_{g,h} \otimes Y_{h,k} = (X^G_{g^{-1}h} \otimes Y^G_{h^{-1}k})_{g,k} \in \mathcal{M}_{g,k},
\end{equation}
where $X^G_a$ denotes a homogeneous object in $\mathscr{F}_a$, and the grading condition gives $\mathscr{F}_{g^{-1}h} \otimes \mathscr{F}_{h^{-1}k} \subset \mathscr{F}_{g^{-1}k}$. 

\bibliographystyle{apsrev4-1}
\bibliography{StringNet}

\end{document}